\documentclass[12pt,english,floatfix,superscriptaddress,aps,prd,preprint,nofootinbib]{revtex4}
\usepackage{amsmath}
\usepackage{amssymb}
\usepackage{amsbsy}
\usepackage[a4paper, margin=1.7cm]{geometry}
\usepackage{amsfonts}
\usepackage{amsopn}
\usepackage{amstext}
\usepackage{graphicx}
\usepackage{amssymb}
\usepackage{amsfonts}
\usepackage{amsmath}
\usepackage{graphicx}
\usepackage[english]{babel}
\usepackage{color}
\usepackage{xcolor}
\usepackage{slashed}
\usepackage{esint}
\usepackage[dvips]{epsfig}
\usepackage[dvips]{graphicx}
\usepackage{float}
\usepackage{units}
\usepackage{textcomp}
\usepackage{placeins}
\usepackage{hyperref}             
\hypersetup{
    colorlinks=true,              
    breaklinks=true,              
    citecolor=blue,               
    linkcolor=[rgb]{0,0.5,0.9},   
    urlcolor=red,                 
    filecolor=green               
}

\usepackage{hyperref}
\usepackage{slashed}

\newcommand{\ie}{\begin{equation}}
\newcommand{\fe}{\end{equation}}
 \newcommand{\bq}{\begin{equation}}
 \newcommand{\eq}{\end{equation}}
 \newcommand{\bqn}{\begin{eqnarray}}
 \newcommand{\eqn}{\end{eqnarray}}

\begin{document}

\title{Thermodynamic and statistical properties of a multifractional modified dispersion relation via the grand-canonical ensemble}


\author{A. A. Ara\'{u}jo Filho}
\email{dilto@fisica.ufc.br}
\affiliation{Departamento de Física, Universidade Federal da Paraíba, Caixa Postal 5008, 58051--970, João Pessoa, Paraíba,  Brazil.}
\affiliation{Departamento de Física, Universidade Federal de Campina Grande Caixa Postal 10071, 58429-900 Campina Grande, Paraíba, Brazil.}
\affiliation{Center for Theoretical Physics, Khazar University, 41 Mehseti Street, Baku, AZ-1096, Azerbaijan.}

\date{\today}

\begin{abstract}

We study the thermodynamic and statistical properties of a gas governed by a multifractional modified dispersion relation of the form $\omega^{2}=k^{2}+4E_{*}^{-1/2}k^{5/2}$, where $E_{*}$ sets the characteristic scale of the multifractional correction. Working within the grand--canonical ensemble, we derive the modified density of states, the grand potential, the partition function, and the main thermodynamic quantities for both bosonic and fermionic sectors. The deformation changes the available phase--space distribution and produces nonstandard thermal scalings controlled by the ratio $T/E_{*}$. In the infrared regime, the usual relativistic gas behavior is recovered with leading corrections proportional to powers of $(T/E_{*})^{1/2}$. In the ultraviolet regime, the density of states scales as $\varrho(\omega)\propto \omega^{7/5}$, corresponding to an effective density--of--states dimension $d_{\mathrm{eff}}=12/5$. As a consequence, the Stefan--Boltzmann law is deformed from $u\propto T^{4}$ to $u\propto E_{*}^{3/5}T^{17/5}$, while the equation--of--state parameter approaches $w=5/12$ instead of the standard radiation value $w=1/3$. We also analyze thermal stability, particle number and energy fluctuations, Bose--Einstein condensation, and the degenerate Fermi gas limit. The multifractional correction increases the critical temperature of a conserved bosonic gas and modifies the Fermi energy, pressure, sound speed, and low--temperature heat capacity of degenerate fermions. {A direct fit to the COBE/FIRAS monopole spectrum yields $E_{*}>0.36\,\mathrm{MeV}$ at $95\%$ CL, while the conservative GWTC--3 propagation constraint gives $E_{*}>1.18\times10^{19}\,\mathrm{GeV}$ at $90\%$ credibility when the dispersion relation is assumed to be universal.}
\end{abstract}


\maketitle
\tableofcontents

\section{Introduction }

The possibility that the relativistic dispersion relation may receive corrections at high energies has been widely discussed in quantum--gravity phenomenology. Such corrections arise in different settings, including massive--graviton models, Lorentz--violating extensions of gravity, deformed relativistic symmetries, noncommutative geometry, Hořava--Lifshitz gravity, extra--dimensional scenarios, effective field theory descriptions of Planck--scale physics, and others \cite{Anacleto:2015awa,Anacleto:2015mma,Will:1998,AmelinoCamelia1998,ColemanGlashow1999,AmelinoCamelia2002,MagueijoSmolin2002,Mattingly2005,Jacobson2006,Horava:2009,Hossenfelder2013,Kostelecky:2003fs,Yunes:2016,Maluf:2014dpa,Delhom:2020gfv}. In most of these approaches, the modified dispersion relation is introduced as a propagation law for elementary or effective excitations. Its physical consequences, however, are not restricted to propagation. Since the energy spectrum determines the map between momentum modes and energy levels, any deformation of this relation changes the density of states and, consequently, the thermal behavior \cite{Anacleto:2018wlj,Araujo:2023izx,Aguirre:2021tpk,Costa:2023siz,Ospedal:2025gpo,Parai:2024dxp,Filho:2021rin,AraujoFilho:2024iox,AraujoFilho:2025fwd,Brito:2015csa}.

In addition, modified dispersion relations have become an important tool in gravitational wave phenomenology \cite{Bailey:2023lzy,li2024power,AraujoFilho:2026oiy,wang2025modified,AraujoFilho:2026vcf,AraujoFilho:2026zyt,mewes2019signals,Amarilo:2023wpn}. The observation of compact--binary coalescences made it possible to constrain deviations from general--relativistic wave propagation, including massive--graviton effects and more general frequency--dependent corrections \cite{Abbott2016PRL,Abbott2016Tests,Yunes:2016,Kostelecky:2016,Arzano:2016,Abbott2019Tests,Abbott2021Tests}. In this context, several parametrizations have been used to connect observational bounds with effective quantum--gravity scales \cite{Arzano:2016,Yunes:2016}. Although many of these studies focus on the propagation of gravitons, the same mathematical dispersion laws may also describe effective quasiparticles or collective modes in other physical systems.

Among the frameworks leading to nonstandard dispersion relations, multifractional spacetime models occupy a special position. They are based on the idea that the effective measure and dimensional properties of spacetime depend on the scale at which the geometry is probed \cite{Calcagni:2010,Calcagni:2012a,Calcagni2013,Calcagni2017}. At macroscopic scales, the ordinary continuum description is recovered, whereas at microscopic scales anomalous scaling may appear. This dimensional flow is not exclusive to multifractional theories. Related forms of dimensional reduction occur in causal dynamical triangulations, asymptotic safety, loop quantum gravity, Hořava--Lifshitz gravity, nonlocal gravity, and other quantum--gravity approaches \cite{Ambjorn2005,LauscherReuter2005,Modesto2009,Horava:2009,Carlip2017,Calcagni:2016}. Multifractional models are useful in the present context because they provide a controlled phenomenological prescription for encoding scale--dependent geometry into the energy spectrum.

On the other hand, the thermodynamic relevance of such a spectrum is immediate. In an ordinary massless relativistic gas, the familiar temperature scalings of the pressure, energy density, entropy density, and heat capacity are tied to the standard density of states. Once the dispersion relation is deformed, the number of modes available at a given energy is redistributed. This mechanism has appeared in several studies of statistical mechanics with modified dispersion relations, including analyses of black body radiation, quantum gases, Bose--Einstein condensation, degenerate fermions, white dwarf phenomenology, generalized uncertainty principle corrections, and doubly special relativity inspired thermodynamics \cite{Camacho2006,Das:2009qb,rajagopal2024unruh,AmelinoCamelia2008,NozariSaghafi2012,MajhiVagenas2013,ChungHassanabadi2019,TawfikDiab2015,furtado2023thermal,rajagopal2024kappa}. 

Furthermore, the grand--canonical ensemble is especially appropriate for studying these effects because it treats thermal and particle number fluctuations within the same framework \cite{Pathria2011,Huang1987,Kardar2007,LandauLifshitz1980}. This is relevant for a deformed gas since the modified spectrum affects both the average thermodynamic quantities and their fluctuations. The ensemble also allows the bosonic and fermionic sectors to be handled in a unified notation. In the bosonic case, a conserved particle number makes it possible to examine how the deformation changes the saturation of excited states and the Bose--Einstein condensation temperature. In the fermionic case, the same spectrum modifies the Fermi surface, the zero temperature equation of state, the sound speed, and the low--temperature heat capacity.

The ultraviolet behavior is particularly important. In multifractional scenarios, the high--energy spectrum may no longer reproduce the usual relativistic scaling. This modifies the effective density of thermal states and changes the power laws governing the Stefan--Boltzmann relation and the equation of state. Such behavior is naturally connected with the broader idea of scale--dependent dimensionality in quantum gravity \cite{Calcagni:2010,Calcagni:2012a,Calcagni2017,Carlip2017}. From the thermodynamic viewpoint, the gas behaves as if the distribution of accessible modes were controlled by an effective density of states dimension different from the standard spatial value. {This therefore provides} a direct way to correlate the geometric content of multifractional spacetime into equilibrium statistical mechanics.

{
It is important to emphasize that the modified dispersion relations considered in this work, including the multifractional branch adopted below, violate the standard Lorentz invariance when written in terms of the physical energy and momentum of a given frame. In multifractional theories, this effect is related to the anomalous spacetime measure, which induces scale dependent geometric profiles and modifies the usual relation $E^{2}=p^{2}+m^{2}$. In other words, the scale $E_{*}$ should be {regarded} as a phenomenological parameter. Solar system, laboratory and astrophysical tests already impose strong bounds on preferred-frame and Lorentz--violating effects, while GW170817 and the LIGO--Virgo--KAGRA catalogues severely constrain the speed, dispersion, birefringence and polarization content of gravitational waves \cite{Colladay:1998fq,Kostelecky:2003fs,Kostelecky:2008ts,Bailey:2006fd,Battat:2007uh,Will:2014kxa,AraujoFilho:2024mvz,Mirshekari:2011yq,Abbott:2017xzu,Abbott2021Tests,Niu:2022yhr,Gong:2023ruw,AraujoFilho:2026oiy,AraujoFilho:2026vcf,AraujoFilho:2026zyt,Amarilo:2023wpn}. For multifractional models, the precise bounds depend on the realization of the theory, the dispersion branch, the fractional exponents, the coupled sector and possible logarithmic oscillations of the measure \cite{Calcagni:2017LIV,Calcagni:2021ipd,Addazi:2018uhd}. In the present analysis, we do not assume an observable violation in photons or gravitons at currently tested energies. Instead, the modified dispersion relation is used as an effective spectrum for a thermal gas, with the standard relativistic behavior recovered for $T/E_{*}\ll1$. In this manner, compatibility with present constraints requires the Lorentz--violating correction to be infrared--suppressed, sector--dependent, or associated with multifractional parameters lying beyond the sensitivity of current solar system and gravitational wave tests.
}

In this work, we investigate the thermodynamic and statistical properties of a gas governed by a multifractional modified dispersion relation within the grand--canonical ensemble. The analysis covers the modified density of states, the grand potential, the partition function, the thermodynamic functions, the equation of state, the deformed Stefan--Boltzmann law, the fluctuation sector, Bose--Einstein condensation, and the degenerate Fermi gas limit. The infrared regime recovers the standard massless relativistic gas with corrections controlled by the multifractional scale, while the ultraviolet regime displays nonstandard thermal powers, a modified effective density of states, and an equation of state parameter different from the usual radiation value.


\section{Multifractional modified dispersion relation }
\label{sec:multifractional_mdr}

{
Before specifying the dispersion relation, let us clarify which multifractional realization is considered in the present work. Multifractional theories are not unique; they include, for instance, the theory with ordinary derivatives $T_1$, the theory with weighted derivatives $T_v$, the theory with $q$-derivatives $T_q$, and other formulations involving fractional differential operators, as discussed in Ref.~\cite{Calcagni:2016} and in the updated review~\cite{Calcagni:2021ipd}. In this paper, we do not consider the whole class of multifractional models. Instead, we restrict ourselves to the effective modified-dispersion-relation sector motivated by the multifractional theory with $q$-derivatives, $T_q$. In this framework, the geometric coordinates $q^\mu(x^\mu)$ encode the scale-dependent measure structure of spacetime, while the ordinary relativistic behavior is recovered in the infrared limit.

More specifically, we adopt the factorizable multifractional measure and neglect possible log-periodic oscillations of the measure, working with the leading zero-mode/binomial profile. The corresponding one-particle spectrum is then treated phenomenologically as an effective dispersion relation for the thermal excitations. In other words, the analysis below should be understood as a grand-canonical study of a gas governed by the timelike branch of the $T_q$-motivated multifractional modified dispersion relation.}

{It is important to distinguish the physical momentum variables from the geometric momenta entering the Fourier formulation of the multifractional theory with $q$--derivatives. We denote the physical energy {and} momentum variables in the fractional picture by
$k^\mu=(\omega,\mathbf{k})$,
whereas the geometric momenta conjugate to the geometric coordinates $q^\mu(x^\mu)$ are denoted by
\begin{equation}
\widetilde{p}^{\,\mu}(k^\mu) = \frac{1}{q^\mu(1/k^\mu)}.
\end{equation}
In a generic multifractional configuration, the momentum space measure written in terms of the physical variables may therefore contain the Jacobian associated with the map $k^\mu\mapsto\widetilde{p}^{\,\mu}(k^\mu)$.

However, the timelike configuration considered in the present work has a nontrivial geometric profile only in the temporal direction, while its spatial profiles are trivial:
\begin{equation}
\nonumber
q^i(x^i)=x^i, \qquad i=1,2,3.
\end{equation}
Therefore,
\begin{equation}
\nonumber
\widetilde{p}^{\,i}(k^i)=k^i, \qquad \mathrm{d}^3\widetilde{\mathbf{p}}(\mathbf{k})=\mathrm{d}^3\mathbf{k}
\end{equation}
In this way, no additional measure weight appears in the spatial momentum-space integrals considered below. The multifractional deformation is instead encoded in the temporal sector and in the resulting physical on--shell relation $\omega=\omega(k)$, with $k=|\mathbf{k}|$ denoting the physical spatial momentum.}

Several approaches to quantum gravity and effective spacetime models suggest that the relativistic dispersion relation may be modified at high energies or short distances. In such scenarios, the propagation of some systems can acquire corrections controlled by a characteristic scale associated with the underlying microscopic structure of spacetime. {Within this effective $T_q$-motivated description, the modified dispersion relation can be parametrized as}
\begin{equation}
E^2 = ( {k} c)^2 + A({k}c)^\alpha ,
\label{generic_mdr}
\end{equation}
where $A$ measures the strength of the correction and $\alpha$ determines the energy dependence of the modification \cite{Will:1998,Yunes:2016,Arzano:2016}. The standard relativistic result is recovered when $A\rightarrow0$.

The parametrization \eqref{generic_mdr} is general enough to describe several modified--propagation mechanisms. For example, massive--graviton models correspond to $\alpha=0$, while other values of $\alpha$ can arise in Lorentz--violating gravity, extra--dimensional models, Ho\v{r}ava--Lifshitz gravity, and quantum spacetime scenarios \cite{Will:1998,Horava:2009,Kostelecky:2016,Yunes:2016}. In this work, we focus on the case motivated by multifractional spacetime.

Multifractional spacetime models are based on the possibility that the effective dimensionality and measure structure of spacetime change with the scale at which the geometry is probed \cite{Calcagni:2010,Calcagni:2012a,Calcagni:2016}. At macroscopic scales, the usual smooth spacetime description is recovered. At shorter scales, however, anomalous scaling properties can modify the propagation of fields. 

In the multifractional scenario, two branches are usually distinguished. The timelike branch is characterized by
\begin{equation}
A_{\rm t} = \frac{2E_*^{2-\alpha}}{3-\alpha},
\label{At_multifractional}
\end{equation}
where $E_*$ denotes the characteristic energy scale associated with the onset of multifractional effects. The exponent lies in the interval $2<\alpha<3$, with $\alpha=5/2$ commonly adopted as a representative value
\cite{Calcagni:2010,Calcagni:2012a,Calcagni:2016,Arzano:2016}.

{It is worth clarifying that the choice $\alpha=5/2$ is not imposed by the multifractional construction. In the notation commonly adopted in the multifractional literature, the exponent appearing in the modified dispersion relation is related to the temporal fractional exponent
$\alpha_{0}$ through $ \alpha=3-\alpha_{0}$,  $0<\alpha_{0}<1$. Then, $\alpha=5/2$ corresponds to the representative value $\alpha_{0}=1/2$. In the present work, we adopt this value from the beginning in order to obtain explicit analytical expressions for the grand--canonical integrals and the associated thermodynamic quantities. A systematic treatment for arbitrary $\alpha_{0}$, although certainly interesting, would require a separate and considerably more general analysis.}

In addition, the restriction to the timelike branch deserves some clarification. In this branch, the multifractional correction enters the dispersion relation with a positive sign. For the representative value $\alpha=5/2$, this leads to $$\omega^2=k^2+4E_*^{-1/2}k^{5/2}$$
which is real and monotonic for all $k>0$. This property is important for the thermodynamic construction below, since the density of states can be obtained from a single-valued map between $k$ and $\omega$.

The spacelike branch, on the other hand, changes the sign of the correction. In the same representative case, one would have
$$\omega_s^2=k^2-4E_*^{-1/2}k^{5/2}
=k^2\left(1-4\sqrt{\frac{k}{E_*}}\right).$$
In other words, the energy is real only for $k\leq E_*/16$. In addition, the corresponding group velocity,
$$v_s(k)=\frac{\mathrm{d}\omega_s}{\mathrm{d}k}
=\frac{1-5\sqrt{k/E_*}}{\sqrt{1-4\sqrt{k/E_*}}},$$
vanishes at $k=E_*/25$ and changes sign before the endpoint of the allowed momentum interval. Therefore, the function $\omega_s(k)$ is not globally monotonic, and the density of states cannot be written using the same single-branch expression adopted in the timelike case. The spacelike branch would require a separate treatment of the finite momentum domain, the different branches of $k(\omega)$, and the behavior near the point where $v_s(k)=0$. Accordingly, we focus here on the timelike branch, while the spacelike branch is left for a separate investigation.

In the present analysis, we restrict attention to the timelike branch. In this manner, using Eq.~\eqref{At_multifractional}, the modified dispersion relation takes the form \cite{Calcagni:2017LIV,Calcagni:2013relpart,Calcagni:2013ku,Calcagni:2016,Calcagni:2016bnn,Calcagni2012b}
\begin{equation}
E^2 = ({k}c)^2 + \frac{2E_*^{2-\alpha}}{3-\alpha}
({k}c)^\alpha .
\label{timelike_mdr_general}
\end{equation}
For the representative multifractional value $\alpha=5/2$, we obtain $A_{\rm t}=4E_*^{-1/2}$,
and therefore
\begin{equation}
E^2 = ({k}c)^2 + 4E_*^{-1/2}({k}c)^{5/2}.
\label{timelike_mdr_alpha_5_2}
\end{equation}
In units $\hbar=c=1$, this becomes
\begin{equation}
\omega^2 = k^2 + 4E_*^{-1/2}k^{5/2}.
\label{timelike_mdr_natural_units}
\end{equation}
On the basis of the above equation, we investigate several aspects of ensemble theory. For this purpose, we adopt the grand--canonical ensemble, which provides a convenient framework for {carrying out the thermodynamic and statistical analysis.}


\section{Grand-canonical framework }
\label{sec:grand_canonical_framework}

{The first part of this section briefly reviews the standard grand--canonical formalism for a homogeneous quantum gas. These relations are recalled for completeness and are not specific to multifractional spacetime, as we must naturally expect. The model--dependent ingredients are the one--particle spectrum $\omega(k)$ and the momentum space measure. As clarified in the previous section, the timelike configuration considered here has a trivial spatial measure, so that $\mathrm{d}^3\widetilde{\mathbf{p}}(\mathbf{k})=\mathrm{d}^3\mathbf{k}$.}

We now formulate the statistical description of a gas of excitations governed by the multifractional modified dispersion relation shown in Eq. (\ref{timelike_mdr_natural_units}). Here, $\omega(k)$ is the energy of a mode with momentum modulus $k=|\mathbf{k}|$, and $E_{*}$ is the characteristic energy scale controlling the multifractional correction. We work in natural units, $\hbar=c=k_{B}=1$, so that temperature, energy, momentum, and inverse length have the same dimension. Equivalently, Eq.~\eqref{timelike_mdr_natural_units} can be written as
\begin{equation}
\omega(k) = k \left( 1+4\sqrt{\frac{k}{E_{*}}} \right)^{1/2}.
\label{eq:omega_k_gc}
\end{equation}
In the limit $E_{*}\rightarrow\infty$, the standard massless relativistic
dispersion relation, $\omega(k)=k$, is recovered.

The grand--canonical ensemble is suitable for systems in which both the energy and the particle number may fluctuate. It also allows bosonic and fermionic gases to be treated within the same formal structure
\cite{PathriaBeale2011,Huang1987,Kardar2007,LandauLifshitz1980}. In this manner, the grand--canonical partition function is defined by
\begin{equation}
\mathcal{Z}_{G} = \mathrm{Tr} \left[ \exp\left( -\beta(\hat{H}-\mu\hat{N}) \right) \right],
\qquad
\beta =\frac{1}{T}.
\label{eq:grand_partition_function}
\end{equation}
In this expression, $\mathcal{Z}_{G}$ is the grand--canonical partition function, $\beta$ is the inverse temperature, $T$ is the temperature, $\hat{H}$ is the Hamiltonian operator, $\hat{N}$ is the particle number operator, and $\mu$ is the chemical potential. The corresponding grand potential is
\begin{equation}
\Phi_{G} = -T\ln\mathcal{Z}_{G},
\label{eq:grand_potential_definition}
\end{equation}
where $\Phi_{G}$ denotes the thermodynamic potential associated with the grand--canonical ensemble.

For a homogeneous gas, the sum over momentum modes is replaced by
\begin{equation}
\sum_{\mathbf{k}} \longrightarrow
\frac{gV}{(2\pi)^{3}} \int \mathrm{d}^{3}k
= \frac{gV}{2\pi^{2}} \int_{0}^{\infty} k^{2}\,\mathrm{d}k,
\label{eq:momentum_sum_gc}
\end{equation}
where $V$ is the spatial volume occupied by the gas, $g$ is the degeneracy factor, and $\mathrm{d}^{3}k$ is the volume element in momentum space. In addition, for quantum gases obeying Bose--Einstein or Fermi--Dirac statistics, we have
\begin{equation}
\Phi_{G} = \mp \frac{gVT}{2\pi^{2}} \int_{0}^{\infty} k^{2} \ln \left[ 1 \pm z e^{- \beta\omega(k)} \right]\mathrm{d}k,
\label{eq:grand_potential_gc}
\end{equation}
with $ z=e^{\beta\mu} $ being the fugacity quantity. In Eq.~\eqref{eq:grand_potential_gc}, the upper sign refers to fermions and the lower sign refers to bosons. For these latter ones, the chemical potential must satisfy $\mu\leq\omega_{0}$, where $\omega_{0}$ is the lowest energy of the spectrum. Since the present dispersion relation is massless, $\omega_{0}=0$ so that $\mu\leq 0$. For radiation--like systems, such as photons or gravitons, the particle number is not conserved and one usually sets $\mu=0$, and $z=1$.

As it is straightforward to verify, the mean occupation number is written as
\begin{equation}
f_{\pm}(k) = \frac{1} {z^{-1}e^{\beta\omega(k)}\pm1},
\label{eq:occupation_number_gc}
\end{equation}
with $f_{\pm}(k)$ denoting the average occupation number of a mode with momentum $k$; in addition, the total particle number is
\begin{equation}
N = \frac{gV}{2\pi^{2}} \int_{0}^{\infty} \frac{k^{2}}{z^{-1}e^{\beta\omega(k)}\pm1} \,\mathrm{d}k,
\label{eq:particle_number_gc}
\end{equation}
and the number density is $n=N/V$.

The pressure follows from the grand potential as $ P = -\Phi_{G}/V$. Using Eq.~\eqref{eq:grand_potential_gc}, we obtain
\begin{equation}
P = \pm \frac{gT}{2\pi^{2}}
\int_{0}^{\infty} k^{2} \ln \left[ 1 \pm z e^{-\beta\omega(k)} \right]\mathrm{d}k.
\label{eq:pressure_integral_gc}
\end{equation}
The internal energy is given by
\begin{equation}
U = \frac{gV}{2\pi^{2}} \int_{0}^{\infty}
\frac{k^{2}\omega(k)}{z^{-1}e^{\beta\omega(k)}\pm1} \,\mathrm{d}k,
\label{eq:internal_energy_gc}
\end{equation}
where $U$ denotes the mean internal energy of the gas. The corresponding energy density reads $u = U/V$. 

The entropy can be calculated from
\begin{equation}
S = -\left( \frac{\partial\Phi_{G}}{\partial T} \right)_{\mu,V}.
\label{eq:entropy_derivative_gc}
\end{equation}
Equivalently, it can be obtained through the
thermodynamic identity
\begin{equation}
S = \frac{U-\Phi_{G}-\mu N}{T}.
\label{eq:entropy_identity_gc}
\end{equation}
The heat capacity at fixed volume and chemical potential is
\begin{equation}
C_{V,\mu} = T \left( \frac{\partial S}{\partial T} \right)_{V,\mu}.
\label{eq:heat_capacity_gc}
\end{equation}
We can also see such a quantity as being the measurement of the thermal response of the gas under variations of temperature at fixed $V$ and $\mu$. Furthermore, the local thermal stability requires $C_{V,\mu}>0$.

For the kind of ensemble that we have been considering, it also gives direct access to particle number fluctuations, which reads
\begin{equation}
\left\langle (\Delta N)^{2} \right\rangle = T
\left( \frac{\partial N}{\partial\mu}
\right)_{T,V} = z \left( \frac{\partial N}{\partial z} \right)_{T,V}.
\label{eq:number_fluctuation_mu_gc}
\end{equation}
Using Eq.~\eqref{eq:particle_number_gc}, this becomes
\begin{equation}
\left\langle (\Delta N)^{2} \right\rangle =
\frac{gV}{2\pi^{2}} \int_{0}^{\infty} k^{2} f_{\pm}(k) \left[ 1\mp f_{\pm}(k) \right]\mathrm{d}k.
\label{eq:number_fluctuation_integral_gc}
\end{equation}
For fermions, this expression gives $f_{+}(k)[1-f_{+}(k)]$, reflecting Pauli suppression. For bosons, it gives $f_{-}(k)[1+f_{-}(k)]$, reflecting Bose enhancement. We shall be using these results to examine how the multifractional scale $E_{*}$ modifies the thermodynamic response, the equation of state, and the quantum--statistical sectors of the gas in the forthcoming sections.


\section{Modified density of states }
\label{sec:modified_density_states}

{Because the spatial momentum space Jacobian is unity in the timelike configuration, the counting of one--particle modes is performed directly in terms of the physical spatial momentum $k$. The number of states with momentum modulus smaller than $k$ is therefore
\begin{equation}
\mathcal{N}(k) = \frac{gV}{2\pi^2} \int_0^k \kappa^2\,\mathrm{d}\kappa = \frac{gV}{6\pi^2}k^3,
\end{equation}
where $\kappa$ is only a dummy integration variable. No additional momentum space weight is omitted in this expression, since $\mathrm{d}^3\widetilde{\mathbf{p}}(\mathbf{k})=\mathrm{d}^3\mathbf{k}$ for the branch considered here.} Since the energy is a monotonic function of $k$, the same quantity can be
written in energy space as
\begin{equation}
\mathcal{N}(\omega) = \frac{gV}{6\pi^{2}}k^{3}(\omega),
\label{eq:cumulative_states_omega}
\end{equation}
where $k(\omega)$ is obtained implicitly from the modified dispersion relation.

The density of states in energy space is defined by $\varrho(\omega) = \mathrm{d}\mathcal{N}/\mathrm{d}\omega.$
Using Eq.~\eqref{eq:cumulative_states_omega}, we have
\begin{equation}
\varrho(\omega) = \frac{gV}{2\pi^{2}}
k^{2} \frac{\mathrm{d}k}{\mathrm{d}\omega}
= \frac{gV}{2\pi^{2}} \frac{k^{2}}{v_{g}(k)},
\label{eq:dos_general}
\end{equation}
where $v_{g}(k) = \mathrm{d}\omega/\mathrm{d}k$
is the group velocity. {As argued before, since $k$ denotes the physical spatial momentum, rather than the geometric momentum $\widetilde{p}$, the observable group velocity associated with the isotropic spectrum is $v_g(k)=\mathrm{d}\omega(k)/\mathrm{d}k$. For the spectrum considered here, direct differentiation gives
\begin{equation}
\label{veloicguhere}
v_g(k) = \frac{1+5\sqrt{k/E_*}}{\sqrt{1+4\sqrt{k/E_*}}}.
\end{equation}}

\begin{figure}
    \centering
    \includegraphics[scale=0.55]{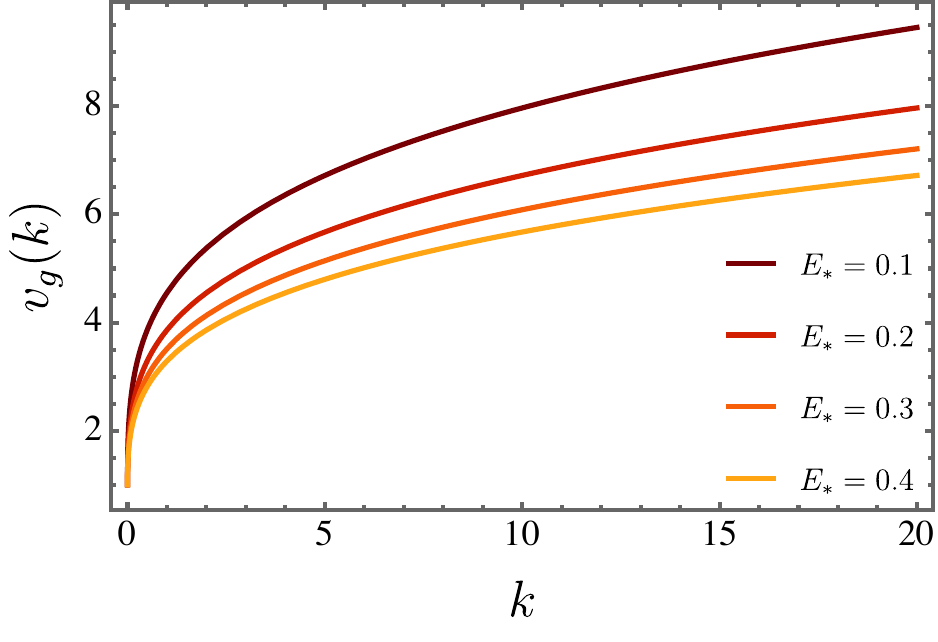}
    \caption{Group velocity $v_{g}(k)$ plotted as a function of $k$ for different values of $E_{*}$.}
    \label{groupvvv}
\end{figure}

In Fig.~\ref{groupvvv}, we display the group velocity $v_{g}(k)$ as a function of $k$ for different values of $E_{*}$. As $E_{*}$ increases, the multifractional correction is suppressed, and $v_{g}(k)$ approaches the standard relativistic value.

In addition, the modified density of states becomes
\begin{equation}
\varrho(\omega) = \frac{gV}{2\pi^{2}} \frac{ k^{2} \left( 1+4\sqrt{k/E_{*}} \right)^{1/2}}
{1+5\sqrt{k/E_{*}}}, \qquad k=k(\omega).
\label{eq:modified_dos}
\end{equation}

\begin{figure}
    \centering
    \includegraphics[scale=0.55]{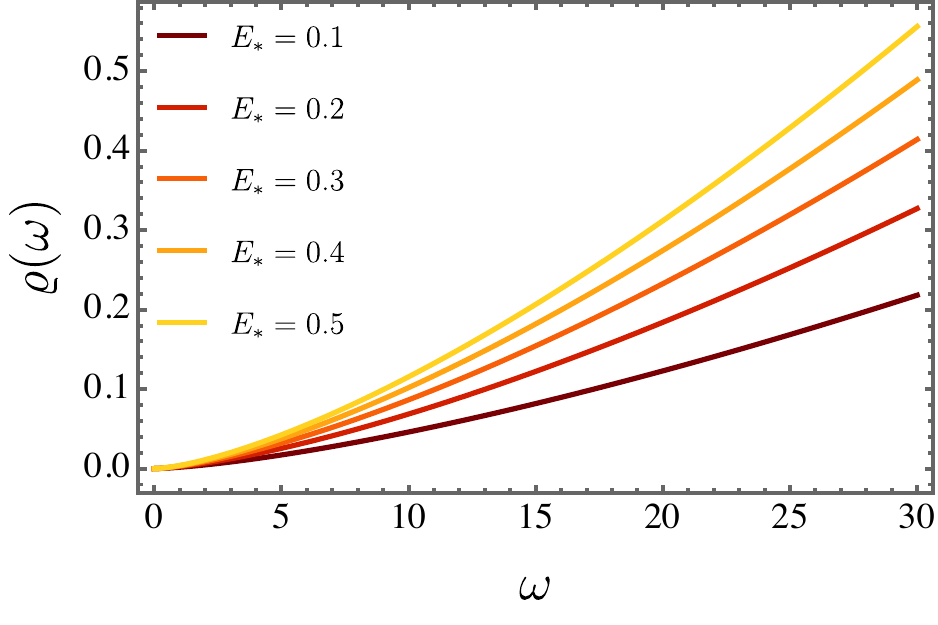}
    \caption{The density of states $\varrho(\omega)$ plotted as a function of $\omega$ for different values of $E_{*}$.  }
    \label{distri}
\end{figure}

In Fig.~\ref{distri}, we display the density of states $\varrho(\omega)$ for different values of $E_{*}$. Unlike the behavior observed for $v_{g}(k)$, $\varrho(\omega)$ increases as $E_{*}$ grows.

In other words, these expressions show that the multifractional correction changes the energy distribution of available states through the factor
\begin{equation}
\mathcal{D}(k;E_{*}) = \frac{ \left( 1+4\sqrt{k/E_{*}} \right)^{1/2}}
{1+5\sqrt{k/E_{*}}}.
\label{eq:dos_deformation_factor}
\end{equation}
Then,
\begin{equation}
\varrho(\omega) = \varrho_{0}(k)\mathcal{D}(k;E_{*}), \qquad \varrho_{0}(k) = \frac{gV}{2\pi^{2}}k^{2}.
\label{eq:dos_factorized}
\end{equation}

In the infrared regime, $k/E_{*}\ll1$, the deformation factor admits the
expansion
\begin{equation}
\mathcal{D}(k;E_{*}) = 1 - 3\sqrt{\frac{k}{E_{*}}} +
\mathcal{O} \left(
\frac{k}{E_{*}} \right).
\label{eq:dos_deformation_ir}
\end{equation}
Consequently,
\begin{equation}
\varrho(\omega) \simeq
\frac{gV}{2\pi^{2}}k^{2} \left[ 1 - 3\sqrt{\frac{k}{E_{*}}} + \mathcal{O}
\left( \frac{k}{E_{*}} \right) \right].
\label{eq:dos_ir_k}
\end{equation}
At leading order, $k\simeq\omega$, and the standard massless result is
recovered:
\begin{equation}
\varrho(\omega) \simeq \frac{gV}{2\pi^{2}}\omega^{2}.
\label{eq:dos_standard_limit}
\end{equation}

The ultraviolet behavior is qualitatively different. For $k/E_{*}\gg1$, the
spectrum behaves as
$\omega(k) \simeq 2E_{*}^{-1/4}k^{5/4}$, 
which implies
\begin{equation}
k(\omega) \simeq \left( \frac{\omega E_{*}^{1/4}}{2} \right)^{4/5}.
\label{eq:k_omega_uv_dos}
\end{equation}
Using $v_{g}(k) \simeq \frac{5}{2}E_{*}^{-1/4}k^{1/4}$,
the density of states becomes
\begin{equation}
\varrho_{\mathrm{UV}}(\omega) \simeq \frac{gV}{5\pi^{2}} E_{*}^{1/4}k^{7/4}(\omega).
\label{eq:dos_uv_k}
\end{equation}
Substituting Eq.~\eqref{eq:k_omega_uv_dos}, one obtains
\begin{equation}
\varrho_{\mathrm{UV}}(\omega) \simeq \frac{gV}{5\pi^{2}} 2^{-7/5} E_{*}^{3/5} \omega^{7/5}.
\label{eq:dos_uv_omega}
\end{equation}
Hence, while the ordinary relativistic gas has
$\varrho(\omega)\propto\omega^{2}$, the multifractional ultraviolet regime
leads to $\varrho_{\mathrm{UV}}(\omega) \propto
\omega^{7/5}$. This result may be expressed in terms of an effective density of states
dimension. If $\varrho(\omega) \propto \omega^{d_{\mathrm{eff}}-1}$, then it gives
$d_{\mathrm{eff}} = 12/5$.
Notice, therefore, that the ultraviolet sector behaves as if the available thermal modes were distributed according to an effective dimension smaller than the usual spatial value $d=3$.


\section{Grand potential and partition function }
\label{sec:grand_potential_partition_function}

To treat the bosonic and fermionic sectors in a unified notation, we introduce the statistical parameter
\begin{equation}
\sigma=
\begin{cases}
+1, & \text{fermions},\\
-1, & \text{bosons}.
\end{cases}
\label{eq:sigma_definition}
\end{equation}
Here, $\sigma$ is only a bookkeeping parameter that distinguishes the two quantum--statistical sectors. In terms of this parameter, the grand--canonical partition function can be written as
\begin{equation}
\mathcal{Z}_{\sigma} = \prod_{\mathbf{k}}
\left[ 1+\sigma z e^{-\beta\omega(k)} \right]^{\sigma},
\label{eq:grand_partition_product}
\end{equation}
where $\prod_{\mathbf{k}}$ denotes the product over all one--particle momentum modes. For $\sigma=+1$, Eq.~\eqref{eq:grand_partition_product} gives the
fermionic product, while for $\sigma=-1$ it gives the inverse bosonic product. Taking the logarithm and passing to the continuum limit, we obtain
\begin{equation}
\ln\mathcal{Z}_{\sigma} =
\sigma \frac{gV}{2\pi^{2}} \int_{0}^{\infty}
k^{2} \ln \left[ 1+\sigma z e^{-\beta\omega(k)} \right] \mathrm{d}k .
\label{eq:log_grand_partition_k}
\end{equation}
The corresponding grand potential is $\Phi_{\sigma} = -T\ln\mathcal{Z}_{\sigma}$.

Using the density of states obtained in the previous section, the logarithm of the partition function can also be expressed as an energy integral:
\begin{equation}
\ln\mathcal{Z}_{\sigma} = \sigma
\int_{0}^{\infty} \varrho(\omega) \ln
\left[ 1+\sigma z e^{-\beta\omega} \right] \mathrm{d}\omega .
\label{eq:log_grand_partition_dos}
\end{equation}
Accordingly,
\begin{equation}
\Phi_{\sigma} = -\sigma T \int_{0}^{\infty}
\varrho(\omega) \ln \left[ 1+\sigma z e^{-\beta\omega} \right] \mathrm{d}\omega .
\label{eq:grand_potential_dos}
\end{equation}
It is worth mentioning that such representations are particularly useful when the infrared and ultraviolet limits are analyzed directly from the scaling behavior of $\varrho(\omega)$.

{Within its convergence domain, the} logarithm appearing in Eq.~\eqref{eq:log_grand_partition_k} admits the series expansion
\begin{equation}
\sigma \ln \left[ 1+\sigma z e^{-\beta\omega(k)}
\right] = \sum_{\ell=1}^{\infty}
\frac{ (-\sigma)^{\ell+1}z^{\ell}}{\ell} e^{ -\ell\beta\omega(k)},
\label{eq:logarithm_expansion}
\end{equation}
with $\ell$ being a positive integer that labels the fugacity expansion.

{
The term--by--term expansion in Eq.~\eqref{eq:logarithm_expansion} requires
\[\left|z e^{-\beta\omega(k)}\right|<1.\]
Since the present spectrum is gapless, with $\omega(0)=0$, this condition is satisfied throughout the complete momentum domain for $0<z<1$. Results at $z=1$ are understood as the boundary limit $z\rightarrow1^{-}$ whenever the resulting momentum integrals remain finite. For a degenerate Fermi gas, one has $\mu>0$ and $z=e^{\beta\mu}>1$, so that the convergence condition fails for modes satisfying $\omega(k)<\mu$. Therefore, the fugacity expansion in Eq.~\eqref{eq:logarithm_expansion} cannot be used term by term in the degenerate regime. As discussed in Sec.~XI, this regime is instead treated directly through the Fermi--Dirac distribution, using the exact momentum integrals in the zero-temperature limit and the Sommerfeld expansion for low but finite temperatures.
}

Substituting it into Eq.~\eqref{eq:log_grand_partition_k} gives
\begin{equation}
\ln\mathcal{Z}_{\sigma} = \frac{gV}{2\pi^{2}}
\sum_{\ell=1}^{\infty} \frac{
(-\sigma)^{\ell+1}z^{\ell}}{\ell} \,\,\mathcal{I}_{\ell}(\beta,E_{*}),
\label{eq:logZ_series_Iell}
\end{equation}
where the auxiliary thermal integral $\mathcal{I}_{\ell}(\beta,E_{*})$ is
defined by
\begin{equation}
\mathcal{I}_{\ell}(\beta,E_{*}) = \int_{0}^{\infty} k^{2} e^{-\ell\beta\omega(k)}
\mathrm{d}k .
\label{eq:Iell_definition}
\end{equation}
Notice that above quantity contains the dependence of the partition function on the multifractional scale through the deformed energy spectrum.

Introducing the dimensionless integration variable $\xi=\beta k$, Eq.~\eqref{eq:Iell_definition} becomes
\begin{equation}
\mathcal{I}_{\ell}(\beta,E_{*})
= T^{3} \int_{0}^{\infty} \xi^{2} \exp \left[ -\ell \xi \left( 1+4\sqrt{\frac{\xi T}{E_{*}}}
\right)^{1/2} \right] \mathrm{d}\xi .
\label{eq:Iell_dimensionless}
\end{equation}

In the infrared regime, $T/E_{*}\ll1$, we have
\begin{equation}
\mathcal{I}_{\ell}(\beta,E_{*}) \simeq \frac{2T^{3}}{\ell^{3}} - \frac{105\sqrt{\pi}}{8} \frac{T^{7/2}}{E_{*}^{1/2}} \frac{1}{\ell^{7/2}} + \mathcal{O} \left( \frac{T^{4}}{E_{*}} \right).
\label{eq:Iell_ir_expansion}
\end{equation}
Substituting this result into Eq.~\eqref{eq:logZ_series_Iell}, we find
\begin{equation}
\frac{\ln\mathcal{Z}_{\sigma}}{V} \simeq \frac{gT^{3}}{\pi^{2}} \mathcal{L}_{4}^{(\sigma)}(z) - \frac{105g}{16\pi^{3/2}} \frac{T^{7/2}}{E_{*}^{1/2}} \mathcal{L}_{9/2}^{(\sigma)}(z) + \mathcal{O} \left( \frac{T^{4}}{E_{*}} \right).
\label{eq:logZ_ir_polylog}
\end{equation}
The function $\mathcal{L}_{s}^{(\sigma)}(z)$ is defined as
\begin{equation}
\mathcal{L}_{s}^{(\sigma)}(z) \equiv \sum_{\ell=1}^{\infty} \frac{ (-\sigma)^{\ell+1}z^{\ell} }{\ell^{s}} = -\sigma\,\mathrm{Li}_{s}(-\sigma z),
\label{eq:statistical_polylog_definition}
\end{equation}
where $\mathrm{Li}_{s}(z)$ denotes the polylogarithm of order $s$. Therefore, for bosons, $ \mathcal{L}_{s}^{(-)}(z) = \mathrm{Li}_{s}(z)$, while for fermions, $\mathcal{L}_{s}^{(+)}(z) = -\mathrm{Li}_{s}(-z)$.

{
The power series representation in Eq.~\eqref{eq:statistical_polylog_definition} is directly obtained within the convergence domain of Eq.~\eqref{eq:logarithm_expansion}, with admissible values at $z=1$ understood as boundary limits. For fermions with $z>1$, the expression $-\mathrm{Li}_{s}(-z)$ represents the analytic continuation of the corresponding Fermi--Dirac function, but it must not be interpreted as the convergent term--by--term fugacity series. In particular, the degenerate Fermi--gas results presented in Sec.~XI are not derived from Eq.~\eqref{eq:logarithm_expansion}.
}

The first term in Eq.~\eqref{eq:logZ_ir_polylog} is the standard massless contribution, while the second term is the leading multifractional correction.

In the ultraviolet regime, the auxiliary integral behaves as
\begin{equation}
\mathcal{I}_{\ell}^{\mathrm{UV}} \simeq \frac{4}{5} 2^{-12/5} \Gamma \left( \frac{12}{5} \right) E_{*}^{3/5} T^{12/5} \ell^{-12/5}.
\label{eq:Iell_uv}
\end{equation}
Here, $\Gamma(s)$ is the Euler gamma function. Inserting
Eq.~\eqref{eq:Iell_uv} into Eq.~\eqref{eq:logZ_series_Iell}, one obtains
\begin{equation}
\frac{\ln\mathcal{Z}_{\sigma}^{\mathrm{UV}}}{V} \simeq \frac{ 2^{-7/5}g } {5\pi^{2}} \Gamma \left( \frac{12}{5} \right) E_{*}^{3/5} T^{12/5} \mathcal{L}_{17/5}^{(\sigma)}(z).
\label{eq:logZ_uv}
\end{equation}
Consequently, the ultraviolet form of the grand potential density is
\begin{equation}
\frac{\Phi_{\sigma}^{\mathrm{UV}}}{V} \simeq - \frac{ 2^{-7/5}g } {5\pi^{2}} \Gamma \left( \frac{12}{5} \right) E_{*}^{3/5} T^{17/5} \mathcal{L}_{17/5}^{(\sigma)}(z).
\label{eq:grand_potential_uv}
\end{equation}

The infrared expression in Eq.~\eqref{eq:logZ_ir_polylog} gives the leading departure from the usual relativistic gas, whereas Eq.~\eqref{eq:grand_potential_uv} displays the high-temperature power law that will determine the modified thermal equation of state. Therefore, notice that with all these preliminaries that we have developed so far, we shall be able to address the thermodynamic behavior in the next section.


\section{Thermodynamic quantities }
\label{sec:thermodynamic_quantities}

The thermodynamic observables follow from the grand--canonical potential through the usual grand--canonical relations. In the compact notation introduced above, the pressure is
\begin{equation}
P_{\sigma} = -\frac{\Phi_{\sigma}}{V} = \sigma\frac{gT}{2\pi^{2}} \int_{0}^{\infty} k^{2} \ln\left[ 1+\sigma z e^{-\beta\omega(k)} \right] \mathrm{d}k .
\label{eq:pressure_sigma}
\end{equation}
The particle number density is obtained from
\begin{equation}
n_{\sigma} = \frac{1}{V}z \left( \frac{\partial\ln\mathcal{Z}_{\sigma}}{\partial z} \right)_{T,V} = \frac{g}{2\pi^{2}} \int_{0}^{\infty} k^{2}f_{\sigma}(k)\,\mathrm{d}k .
\label{eq:number_density_sigma}
\end{equation}
Here, $f_{\sigma}(k)$ denotes the occupation number written with the statistical parameter $\sigma$ defined via  Eq.~\eqref{eq:occupation_number_gc}.

The internal energy density can be written as
\begin{equation}
u_{\sigma} = -\frac{1}{V} \left( \frac{\partial\ln\mathcal{Z}_{\sigma}}{\partial\beta} \right)_{z,V} = \frac{g}{2\pi^{2}} \int_{0}^{\infty} k^{2}\omega(k)f_{\sigma}(k)\,\mathrm{d}k.
\label{eq:energy_density_sigma}
\end{equation}
The entropy density, denoted by $s_{\sigma}=S_{\sigma}/V$, is obtained from the
grand--canonical identity
\begin{equation}
s_{\sigma} = \frac{u_{\sigma}+P_{\sigma}-\mu n_{\sigma}}{T}.
\label{eq:entropy_density_sigma}
\end{equation}
For radiation--like systems, where $\mu=0$, this reduces to
\begin{equation}
s_{\sigma} = \frac{u_{\sigma}+P_{\sigma}}{T}.
\label{eq:entropy_density_mu_zero}
\end{equation}

The heat capacity density at fixed volume and chemical potential is defined as
\begin{equation}
c_{V,\mu}^{(\sigma)} = T \left(\frac{\partial s}{\partial T} \right)_{\mu}.
\label{eq:heat_capacity_density_definition}
\end{equation}
Using Eq.~\eqref{eq:energy_density_sigma}, one obtains
\begin{equation}
c_{V,\mu}^{(\sigma)} = \frac{g}{2\pi^{2}T^{2}} \int_{0}^{\infty} k^{2} \left[\omega(k)-\mu\right]^{2} f_{\sigma}(k) \left[1-\sigma f_{\sigma}(k)\right] \mathrm{d}k .
\label{eq:heat_capacity_density_sigma}
\end{equation}

{
The factor $f_{\sigma}(k)[1-\sigma f_{\sigma}(k)]$ gives the expected Pauli
suppression for fermions and Bose enhancement for bosons. The condition $c_{V,\mu}^{(\sigma)}>0$ is a necessary local stability condition. The fixed-$z$ and fixed-$\mu$ heat capacities are generally different because $z=e^{\mu/T}$. They coincide in the radiation--like sector, for which $\mu=0$ and
$z=1$.}

The number susceptibility is defined as
\begin{equation}
\chi_{T}^{(\sigma)} = \left( \frac{\partial n_{\sigma}}{\partial\mu} \right)_{T} = \frac{g}{2\pi^{2}T} \int_{0}^{\infty} k^{2}f_{\sigma}(k) \left[1-\sigma f_{\sigma}(k)\right] \mathrm{d}k.
\label{eq:number_susceptibility_sigma}
\end{equation}
The particle number fluctuation per unit volume is therefore
\begin{equation}
\frac{ \left\langle(\Delta N)^{2}\right\rangle_{\sigma} }{V} = T\chi_{T}^{(\sigma)}.
\label{eq:number_fluctuation_density_sigma}
\end{equation}

{
To compare the infrared and ultraviolet limits without extrapolating either
approximation beyond its domain of validity, we introduce the dimensionless variables
\begin{equation}
x=\frac{T}{E_{*}}, \qquad y=\frac{k}{E_{*}}, \qquad \varepsilon(y) = \frac{\omega(k)}{E_{*}} = y\sqrt{1+4\sqrt{y}} .
\label{eq:dimensionless_thermal_variables}
\end{equation}
For the bosonic radiation sector, $\sigma=-1$ and $z=1$, we define
\begin{equation}
\overline{P}_{B} = \frac{P_{B}}{E_{*}^{4}}, \qquad \overline{u}_{B} = \frac{u_{B}}{E_{*}^{4}}, \qquad \overline{n}_{B} = \frac{n_{B}}{E_{*}^{3}}, \qquad \overline{s}_{B} = \frac{s_{B}}{E_{*}^{3}}, \qquad \overline{c}_{V,B} = \frac{c_{V,B}}{E_{*}^{3}} .
\label{eq:dimensionless_observables}
\end{equation}
The corresponding exact expressions are
\begin{align}
\overline{P}_{B}^{\mathrm{exact}}(x) &= -\frac{gx}{2\pi^{2}} \int_{0}^{\infty} y^{2} \ln\left[ 1-e^{-\varepsilon(y)/x} \right] \mathrm{d}y ,
\label{eq:pressure_boson_exact_dimensionless}
\\
\overline{n}_{B}^{\mathrm{exact}}(x) &= \frac{g}{2\pi^{2}} \int_{0}^{\infty} \frac{y^{2}} {e^{\varepsilon(y)/x}-1} \mathrm{d}y  ,
\label{eq:number_boson_exact_dimensionless}
\\
\overline{u}_{B}^{\mathrm{exact}}(x) &= \frac{g}{2\pi^{2}} \int_{0}^{\infty} \frac{y^{2}\varepsilon(y)} {e^{\varepsilon(y)/x}-1} \mathrm{d}y,
\label{eq:energy_boson_exact_dimensionless}
\\
\overline{s}_{B}^{\mathrm{exact}}(x) &= \frac{ \overline{u}_{B}^{\mathrm{exact}}(x) + \overline{P}_{B}^{\mathrm{exact}}(x) }{x},
\label{eq:entropy_boson_exact_dimensionless}
\\
\overline{c}_{V,B}^{\mathrm{exact}}(x) &= \frac{g}{2\pi^{2}x^{2}} \int_{0}^{\infty} y^{2}\varepsilon^{2}(y) \frac{e^{\varepsilon(y)/x}} {\left[e^{\varepsilon(y)/x}-1\right]^{2}} \mathrm{d}y .
\label{eq:heat_capacity_boson_exact_dimensionless}
\end{align}
Notice that these integrals provide the complete crossover between the two asymptotic regimes and remain positive throughout the admissible bosonic domain.
}

{
We first consider the infrared limit $x=T/E_{*}\ll1$. It is important to mention that these expressions below are truncated asymptotic expansions and must not be used when the correction becomes comparable to the leading relativistic contribution.
}
Using Eq.~\eqref{eq:logZ_ir_polylog}, the pressure becomes
\begin{equation}
P_{\sigma}^{\mathrm{IR}} \simeq \frac{gT^{4}}{\pi^{2}} \mathcal{L}_{4}^{(\sigma)}(z) - \frac{105g}{16\pi^{3/2}} \frac{T^{9/2}}{E_{*}^{1/2}} \mathcal{L}_{9/2}^{(\sigma)}(z) + \mathcal{O}\left( \frac{T^{5}}{E_{*}} \right).
\label{eq:pressure_ir_sigma}
\end{equation}
Similarly, the number density becomes
\begin{equation}
n_{\sigma}^{\mathrm{IR}} \simeq \frac{gT^{3}}{\pi^{2}} \mathcal{L}_{3}^{(\sigma)}(z) - \frac{105g}{16\pi^{3/2}} \frac{T^{7/2}}{E_{*}^{1/2}} \mathcal{L}_{7/2}^{(\sigma)}(z) + \mathcal{O}\left( \frac{T^{4}}{E_{*}} \right).
\label{eq:number_density_ir_sigma}
\end{equation}
The internal energy density follows from the temperature derivative of
$\ln\mathcal{Z}_{\sigma}$ at fixed fugacity:
\begin{equation}
u_{\sigma}^{\mathrm{IR}} \simeq \frac{3gT^{4}}{\pi^{2}} \mathcal{L}_{4}^{(\sigma)}(z) - \frac{735g}{32\pi^{3/2}} \frac{T^{9/2}}{E_{*}^{1/2}} \mathcal{L}_{9/2}^{(\sigma)}(z) + \mathcal{O}\left( \frac{T^{5}}{E_{*}} \right).
\label{eq:energy_density_ir_sigma}
\end{equation}
Since $\mu=T\ln z$, the entropy density in the same regime can be written as
\begin{align}
s_{\sigma}^{\mathrm{IR}} \simeq\;& \frac{4gT^{3}}{\pi^{2}} \mathcal{L}_{4}^{(\sigma)}(z) - \frac{gT^{3}}{\pi^{2}} \mathcal{L}_{3}^{(\sigma)}(z)\ln z \nonumber\\ &- \frac{945g}{32\pi^{3/2}} \frac{T^{7/2}}{E_{*}^{1/2}} \mathcal{L}_{9/2}^{(\sigma)}(z) + \frac{105g}{16\pi^{3/2}} \frac{T^{7/2}}{E_{*}^{1/2}} \mathcal{L}_{7/2}^{(\sigma)}(z)\ln z + \mathcal{O}\left( \frac{T^{4}}{E_{*}} \right).
\label{eq:entropy_density_ir_sigma}
\end{align}
For $z=1$, the terms proportional to $\ln z$ vanish.

{
In the ultraviolet limit, $x=T/E_{*}\gg1$, the thermally populated modes probe $k/E_{*}\gg1$, and the dispersion relation approaches $\omega(k)\simeq2E_{*}^{-1/4}k^{5/4}$.}
Equation~\eqref{eq:grand_potential_uv} then gives
\begin{equation}
P_{\sigma}^{\mathrm{UV}} \simeq \mathcal{A} E_{*}^{3/5}T^{17/5} \mathcal{L}_{17/5}^{(\sigma)}(z),
\label{eq:pressure_uv_sigma}
\end{equation}
where
\begin{equation}
\mathcal{A} = \frac{2^{-7/5}g}{5\pi^{2}} \Gamma\left( \frac{12}{5} \right).
\label{eq:A_constant_definition}
\end{equation}
The number density is
\begin{equation}
n_{\sigma}^{\mathrm{UV}} \simeq \mathcal{A} E_{*}^{3/5}T^{12/5} \mathcal{L}_{12/5}^{(\sigma)}(z),
\label{eq:number_density_uv_sigma}
\end{equation}
and the internal energy density becomes
\begin{equation}
u_{\sigma}^{\mathrm{UV}} \simeq \frac{12}{5} \mathcal{A} E_{*}^{3/5}T^{17/5} \mathcal{L}_{17/5}^{(\sigma)}(z).
\label{eq:energy_density_uv_sigma}
\end{equation}
The entropy density is
\begin{equation}
s_{\sigma}^{\mathrm{UV}} \simeq \mathcal{A} E_{*}^{3/5}T^{12/5} \left[ \frac{17}{5} \mathcal{L}_{17/5}^{(\sigma)}(z) - \mathcal{L}_{12/5}^{(\sigma)}(z)\ln z \right].
\label{eq:entropy_density_uv_sigma}
\end{equation}
For vanishing chemical potential, this reduces to
\begin{equation}
s_{\sigma}^{\mathrm{UV}} \simeq \frac{17}{5} \mathcal{A} E_{*}^{3/5}T^{12/5} \mathcal{L}_{17/5}^{(\sigma)}(1).
\label{eq:entropy_density_uv_mu_zero}
\end{equation}
For the radiation--like case, the ultraviolet heat capacity density follows from
Eq.~\eqref{eq:energy_density_uv_sigma} at fixed $z=1$:
\begin{equation}
c_{V}^{(\sigma),\mathrm{UV}} \simeq \frac{204}{25} \mathcal{A} E_{*}^{3/5}T^{12/5} \mathcal{L}_{17/5}^{(\sigma)}(1).
\label{eq:heat_capacity_uv_mu_zero}
\end{equation}

\begin{figure}
    \centering
    \includegraphics[scale=0.435]{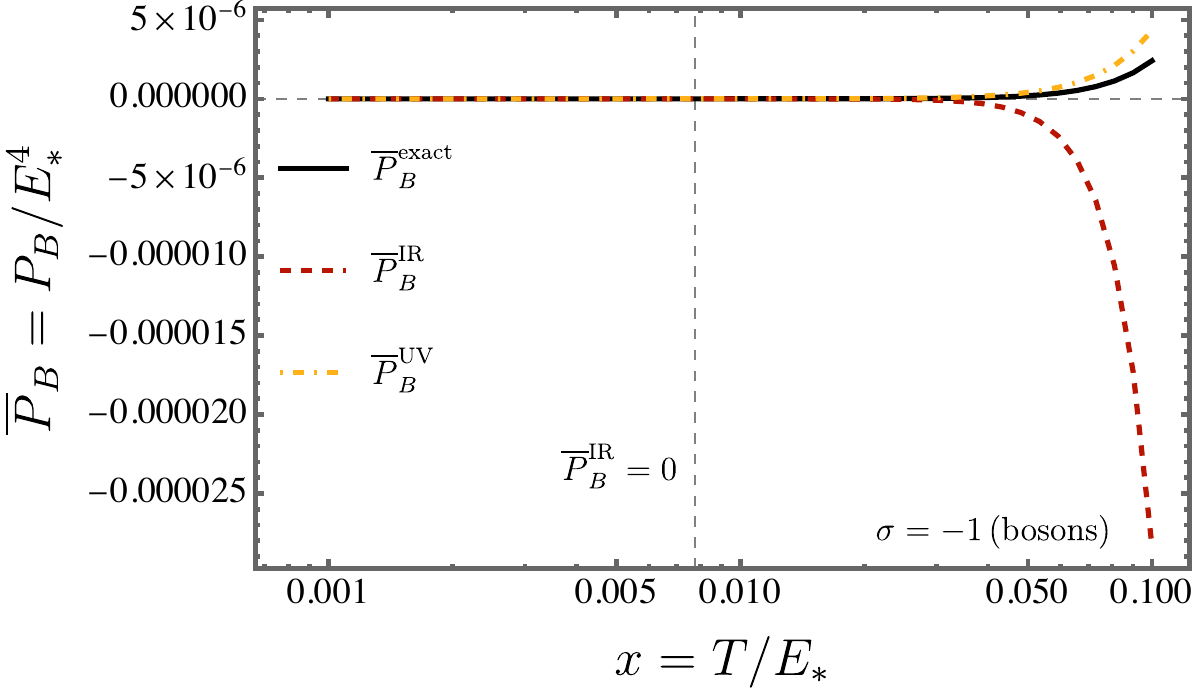}
    \includegraphics[scale=0.435]{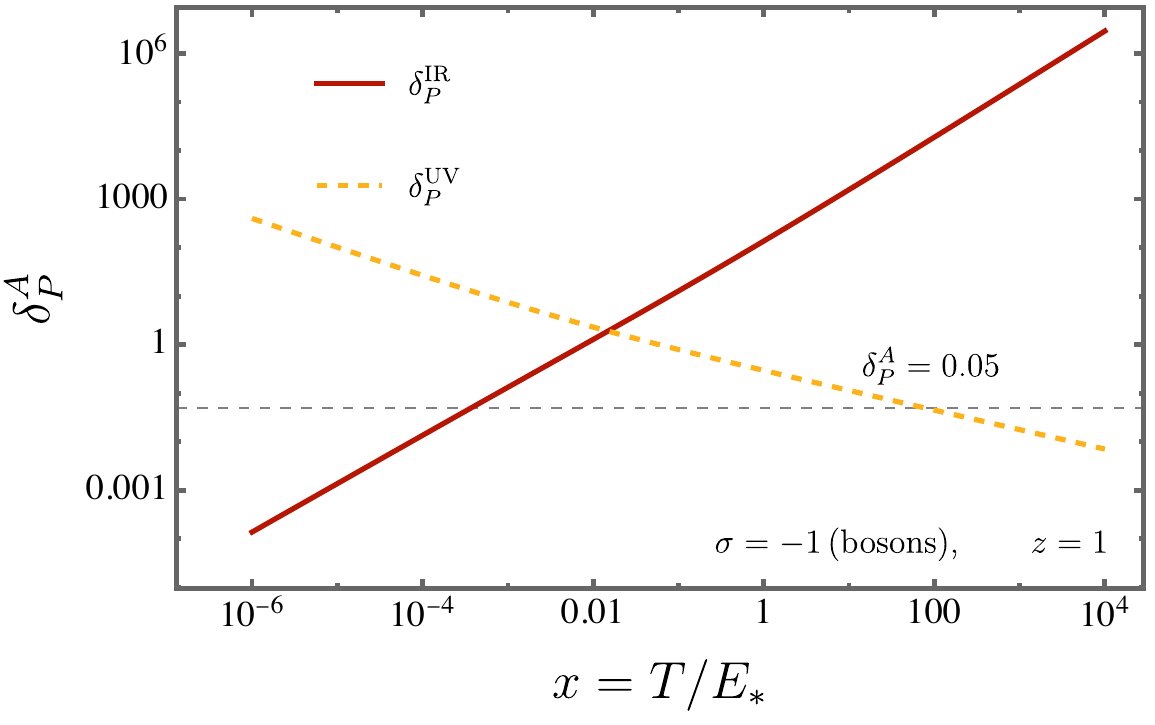}
    \caption{ {Dimensionless bosonic pressure $\overline{P}_{B}=P_{B}/E_{*}^{4}$ as a function of $x=T/E_{*}$ for $\sigma=-1$ and $z=1$. The left panel compares the exact numerical result with the infrared and ultraviolet asymptotic approximations, also indicating the point at which the truncated infrared expression satisfies $\overline{P}_{B}^{\mathrm{IR}}=0$. The right panel displays the corresponding relative deviations $\delta_{P}^{A}$, with $A\in\{\mathrm{IR},\mathrm{UV}\}$, from the exact pressure. The horizontal line $\delta_{P}^{A}=0.05$ represents the $5\%$ accuracy threshold used to determine the validity domain of each asymptotic approximation.}}
    \label{pressureboson}
\end{figure}

\begin{figure}
    \centering
    \includegraphics[scale=0.43]{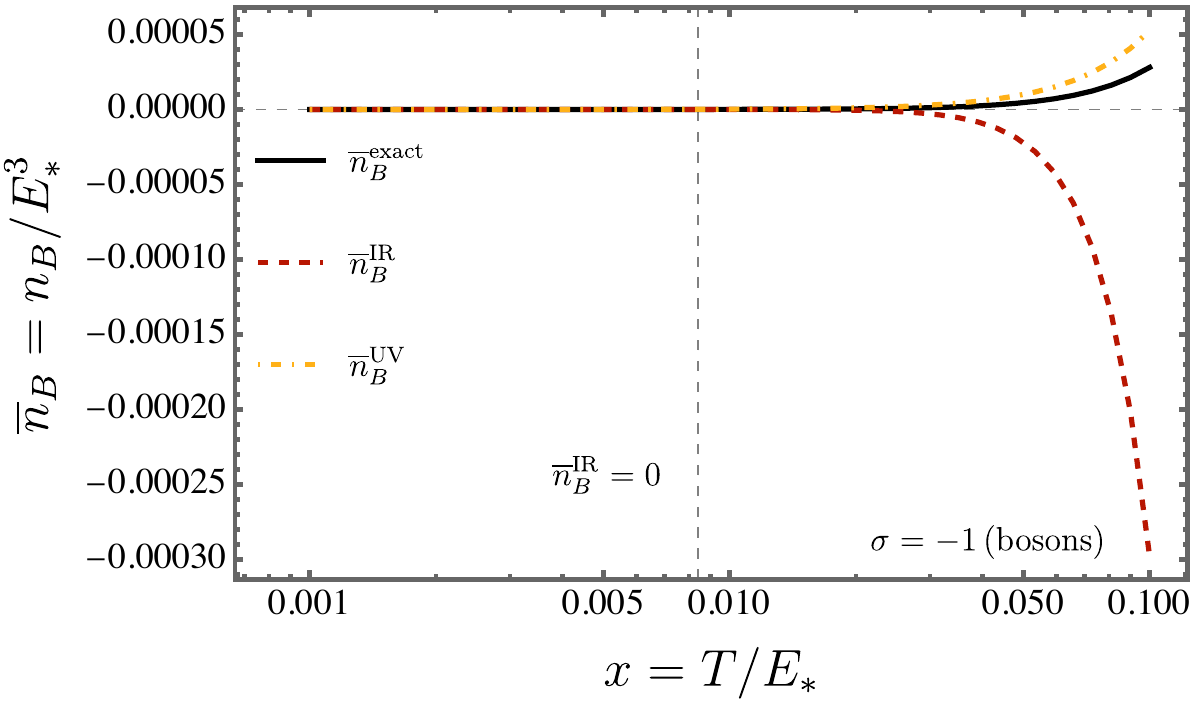}
    \includegraphics[scale=0.43]{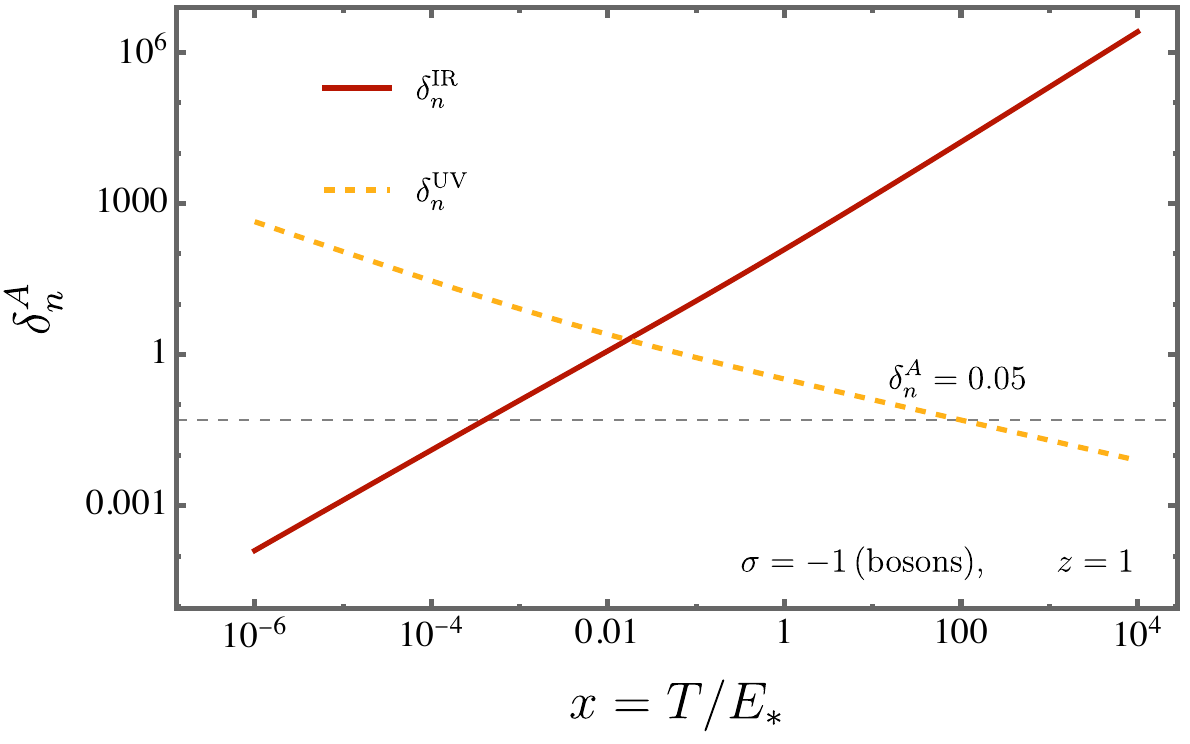}
    \caption{Dimensionless bosonic particle number density $\overline{n}_{B}=n_{B}/E_{*}^{3}$ as a function of $x=T/E_{*}$ for $\sigma=-1$ and $z=1$. The left panel compares the exact numerical result with the infrared and ultraviolet asymptotic approximations and indicates the point at which the truncated infrared expression satisfies $\overline{n}_{B}^{\mathrm{IR}}=0$. The right panel displays the corresponding relative deviations $\delta_{n}^{A}$, with $A\in\{\mathrm{IR},\mathrm{UV}\}$, from the exact particle number density. The horizontal line $\delta_{n}^{A}=0.05$ denotes the $5\%$ accuracy threshold used to determine the validity domain of each asymptotic approximation.
}
    \label{particleboson}
\end{figure}

\begin{figure}
    \centering
    \includegraphics[scale=0.42]{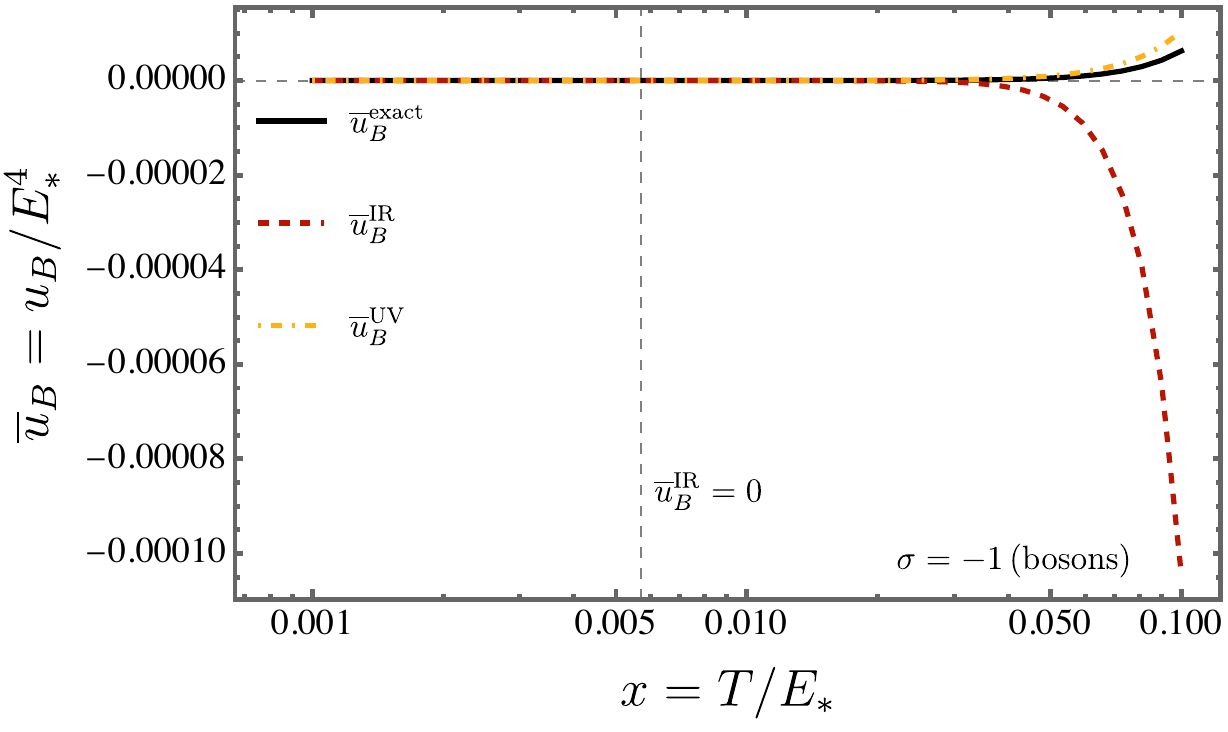}
    \includegraphics[scale=0.42]{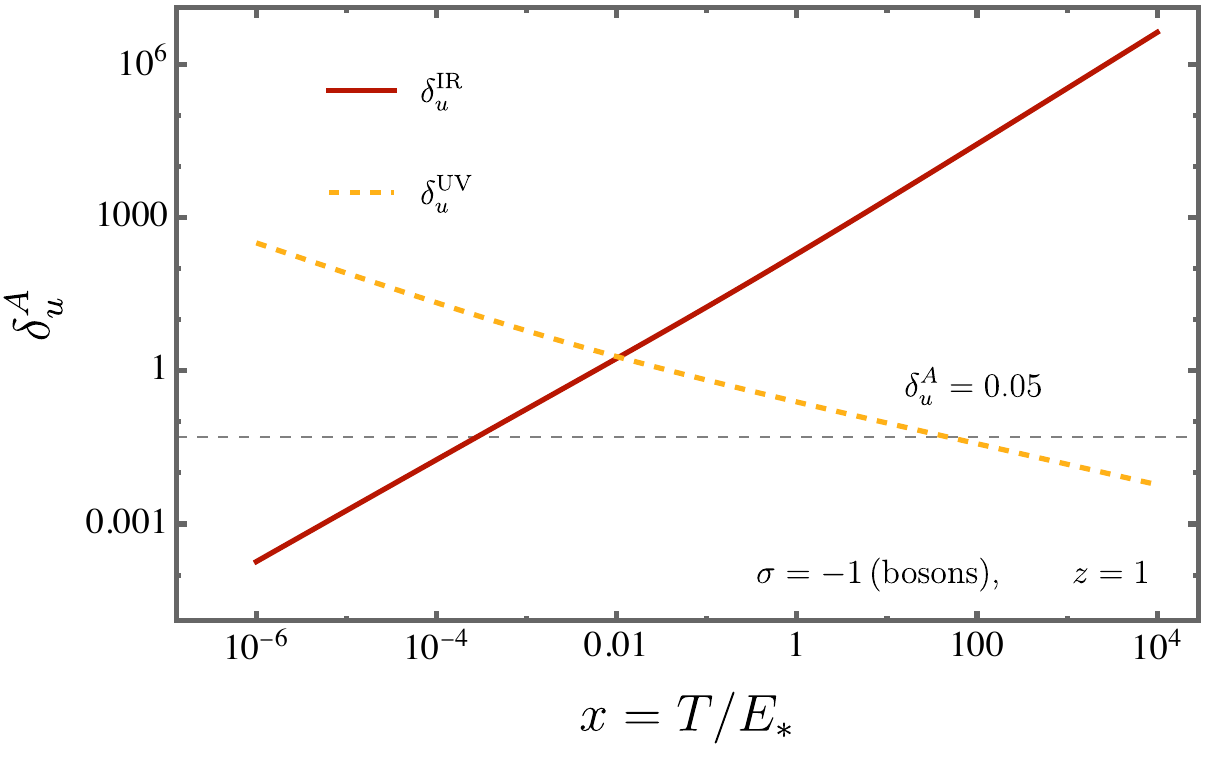}
    \caption{Dimensionless bosonic internal energy density
$\overline{u}_{B}=u_{B}/E_{*}^{4}$ as a function of
$x=T/E_{*}$ for $\sigma=-1$ and $z=1$. The left panel
compares the exact numerical result with the infrared and
ultraviolet asymptotic approximations and indicates the point at
which the truncated infrared expression satisfies
$\overline{u}_{B}^{\mathrm{IR}}=0$. The right panel displays the
corresponding relative deviations $\delta_{u}^{A}$, with
$A\in\{\mathrm{IR},\mathrm{UV}\}$, from the exact internal energy
density. The horizontal line $\delta_{u}^{A}=0.05$ represents
the $5\%$ accuracy threshold used to determine the validity domain
of each asymptotic approximation.
}
    \label{energybosons}
\end{figure}

\begin{figure}
    \centering
    \includegraphics[scale=0.4]{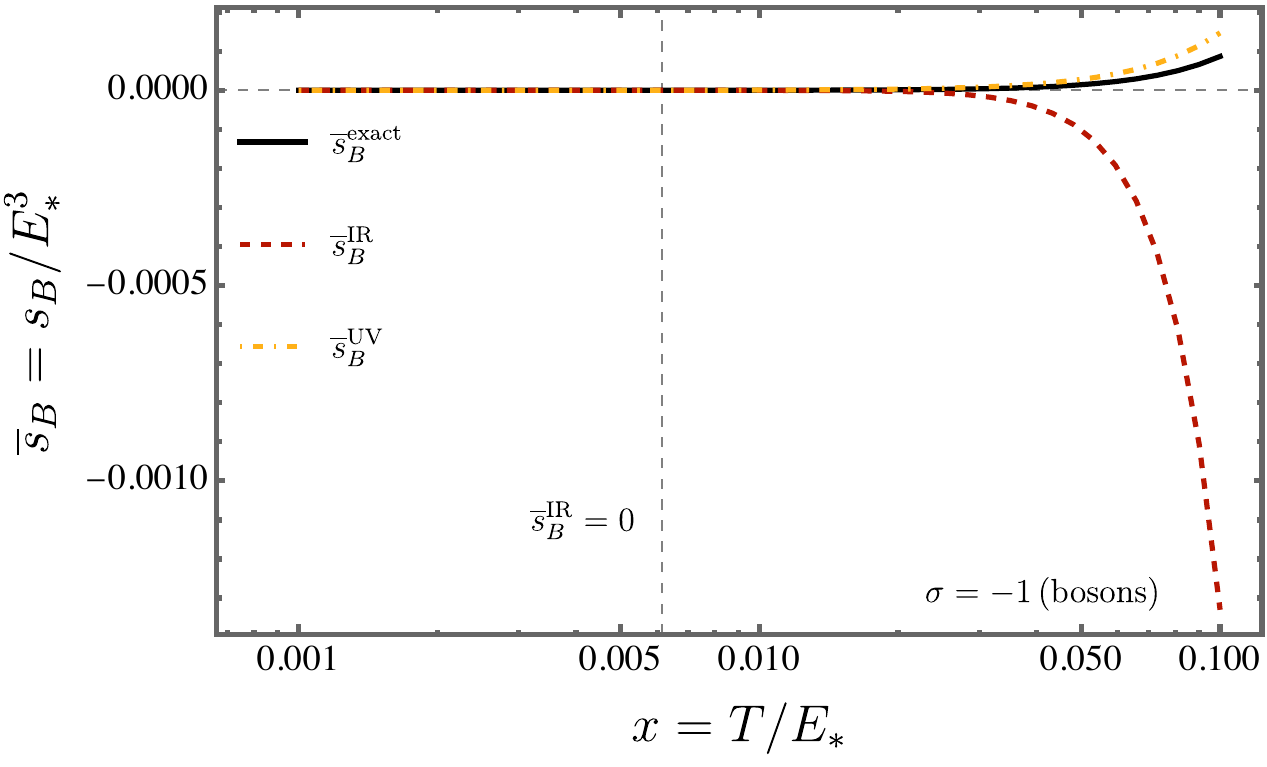}
    \includegraphics[scale=0.4]{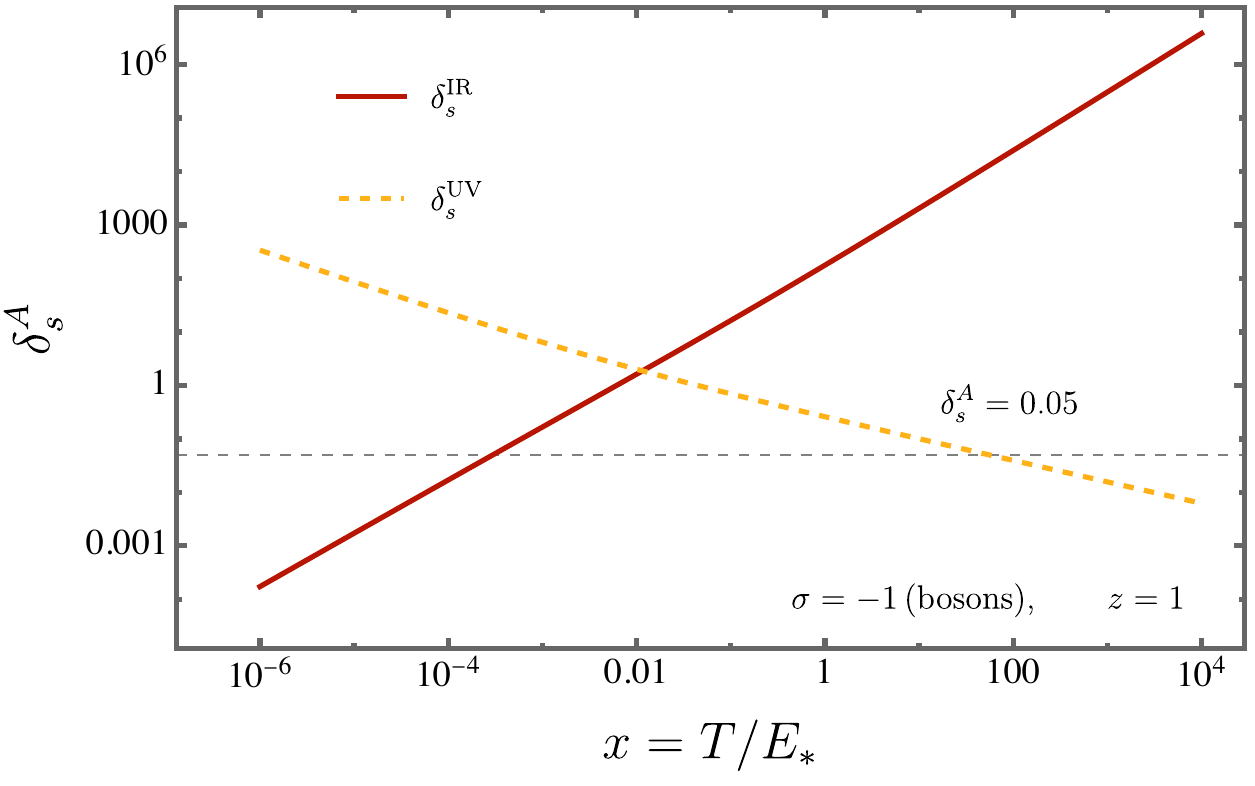}
   \caption{Dimensionless bosonic entropy density
$\overline{s}_{B}=s_{B}/E_{*}^{3}$ as a function of
$x=T/E_{*}$ for $\sigma=-1$ and $z=1$. The left panel
compares the exact numerical result with the infrared and
ultraviolet asymptotic approximations and indicates the point at
which the truncated infrared expression satisfies
$\overline{s}_{B}^{\mathrm{IR}}=0$. The right panel displays the
corresponding relative deviations $\delta_{s}^{A}$, with
$A\in\{\mathrm{IR},\mathrm{UV}\}$, from the exact entropy density.
The horizontal line $\delta_{s}^{A}=0.05$ denotes the $5\%$
accuracy threshold used to determine the validity domain of each
asymptotic approximation.}
    \label{entropybosons}
\end{figure}

\begin{figure}
    \centering
    \includegraphics[scale=0.41]{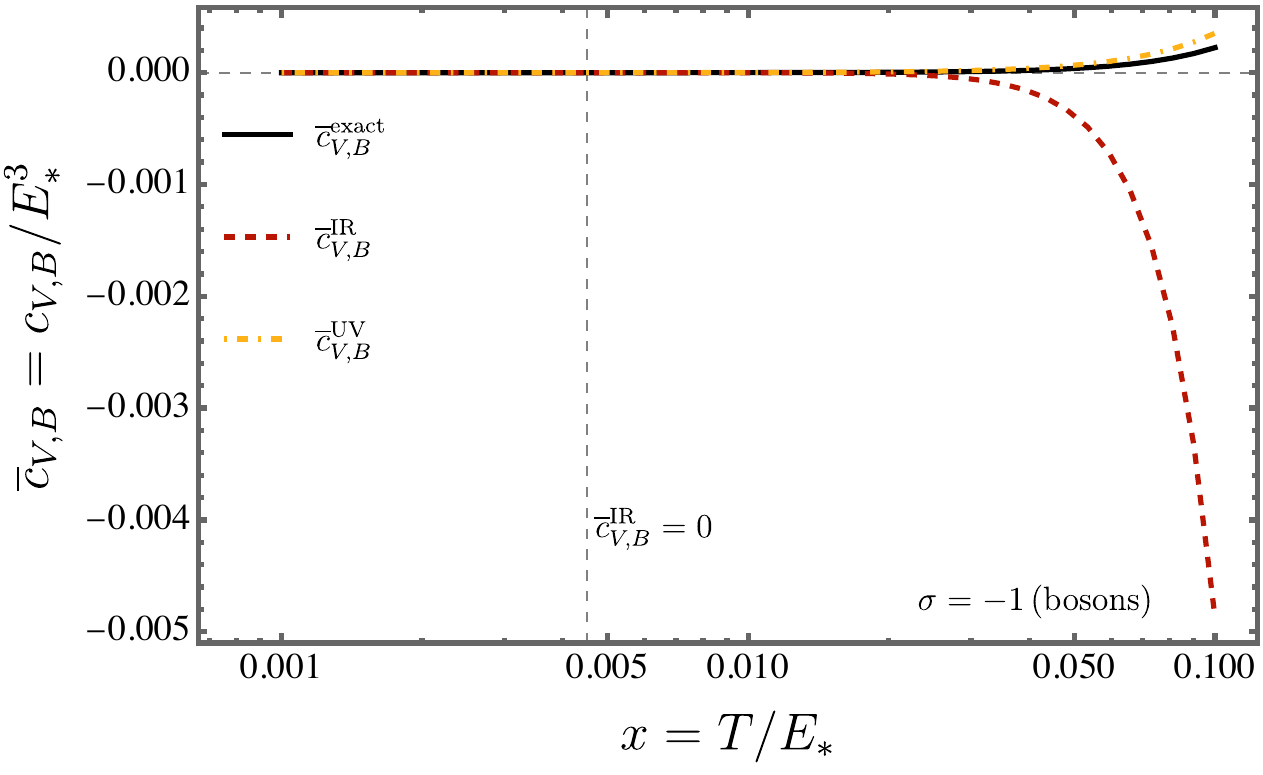}
    \includegraphics[scale=0.41]{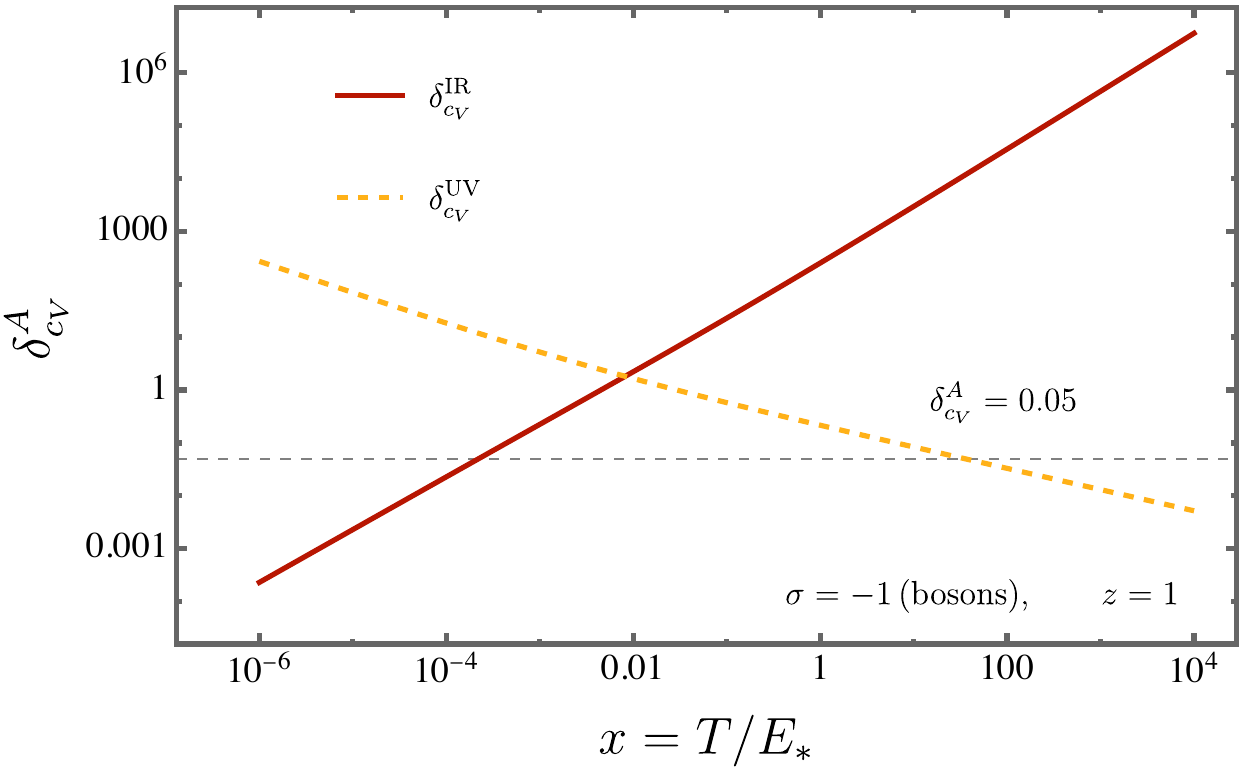}
   \caption{Dimensionless bosonic heat-capacity density
$\overline{c}_{V,B}=c_{V,B}/E_{*}^{3}$ as a function of
$x=T/E_{*}$ for $\sigma=-1$ and $z=1$. The left panel
compares the exact numerical result with the infrared and
ultraviolet asymptotic approximations and indicates the point at
which the truncated infrared expression satisfies
$\overline{c}_{V,B}^{\mathrm{IR}}=0$. The right panel displays the
corresponding relative deviations $\delta_{c_{V}}^{A}$, with
$A\in\{\mathrm{IR},\mathrm{UV}\}$, from the exact heat-capacity
density. The horizontal line $\delta_{c_{V}}^{A}=0.05$ represents
the $5\%$ accuracy threshold used to determine the validity domain
of each asymptotic approximation.}
    \label{heatcapcity}
\end{figure}

{
The positivity of Eq.~\eqref{eq:heat_capacity_uv_mu_zero} shows that the radiation--like ultraviolet sector is locally stable. Equations \eqref{eq:pressure_ir_sigma}--\eqref{eq:heat_capacity_uv_mu_zero} describe two distinct asymptotic limits; they do not provide an interpolation between them.

For an observable $\overline{Q}\in \{\overline{P}_{B},\overline{n}_{B},\overline{u}_{B}, \overline{s}_{B},\overline{c}_{V,B}\}$, we quantify the validity of each approximation by
\begin{equation}
\delta_{\overline{Q}}^{\,\mathrm{A}}(x) = \left| \frac{ \overline{Q}^{\,\mathrm{A}}(x) - \overline{Q}^{\,\mathrm{exact}}(x) }{ \overline{Q}^{\,\mathrm{exact}}(x) } \right|, \qquad \mathrm{A}\in\{\mathrm{IR},\mathrm{UV}\}.
\label{eq:relative_asymptotic_error}
\end{equation}
In Figs.~\ref{pressureboson}--\ref{heatcapcity}, each asymptotic curve is shown only where
\begin{equation}
\delta_{\overline{Q}}^{\,\mathrm{A}}(x)\leq0.05.
\label{eq:five_percent_validity}
\end{equation}

{In particular, in Fig.~\ref{pressureboson}, we compare the exact dimensionless bosonic pressure with its infrared and ultraviolet asymptotic approximations. The left panel shows that the infrared expression reproduces the exact result only for sufficiently small values of $x=T/E_{*}$. As $x$ increases, the truncated infrared expansion departs progressively from the exact curve and eventually becomes negative. This behavior does not indicate a physical negative pressure, but rather the breakdown of the infrared approximation outside its domain of validity. In contrast, the ultraviolet expression becomes increasingly accurate at large $x$, consistently with the high--temperature scaling $\overline{P}_{B}^{\rm UV}\propto x^{17/5}$. The right panel makes these validity domains more explicit through the relative deviations $\delta_{P}^{\rm IR}$ and $\delta_{P}^{\rm UV}$. While the infrared error increases with $x$, the ultraviolet error decreases, and their intersections with the threshold $\delta_{P}^{A}=0.05$ determine the regions in which each asymptotic expression approximates the exact pressure with an accuracy better than $5\%$.}

{
The remaining thermodynamic quantities exhibit the same crossover structure, although with different ultraviolet powers. In Fig.~\ref{particleboson}, the infrared approximation for the particle number density agrees with the exact result at small $x$, whereas the ultraviolet expression approaches it at large $x$ with $\overline{n}_{B}^{\rm UV}\propto x^{12/5}$. Figure~\ref{energybosons} shows that the internal energy density follows the ultraviolet scaling $\overline{u}_{B}^{\rm UV}\propto x^{17/5}$. In Fig.~\ref{entropybosons}, the exact entropy density remains positive, while the negative continuation of the infrared curve only signals the breakdown of the truncated expansion; the ultraviolet behavior is $\overline{s}_{B}^{\rm UV}\propto x^{12/5}$. Finally, Fig.~\ref{heatcapcity} shows that the exact heat capacity remains positive throughout the considered domain, confirming local thermal stability, while its ultraviolet limit scales as $\overline{c}_{V,B}^{\rm UV}\propto x^{12/5}$. In all these cases, the right panels identify the corresponding $5\%$ validity domains through the intersections of $\delta_{Q}^{\rm IR}$ and $\delta_{Q}^{\rm UV}$ with $\delta_{Q}^{A}=0.05$.
}

}

\section{Modified equation of state }
\label{sec:modified_equation_of_state}

The equation of state can be obtained by comparing the pressure with the internal energy density. For this purpose, it is useful to start from the kinetic representation of the pressure,
\begin{equation}
P_{\sigma} = \frac{g}{6\pi^{2}}
\int_{0}^{\infty} k^{3} \frac{\mathrm{d}\omega}{\mathrm{d}k} f_{\sigma}(k) \,\mathrm{d}k .
\label{eq:kinetic_pressure_sigma}
\end{equation}
Combining this expression with the energy density, we have
\begin{equation}
P_{\sigma} = \frac{g}{2\pi^{2}}
\int_{0}^{\infty} k^{2}\omega(k) \mathcal{W}(k;E_{*}) f_{\sigma}(k) \,\mathrm{d}k ,
\label{eq:pressure_weighted_form}
\end{equation}
where the momentum dependent equation of state weight is defined as
\begin{equation}
\mathcal{W}(k;E_{*}) = \frac{1}{3} \bigg( \frac{k}{\omega(k)} \frac{\mathrm{d}\omega}{\mathrm{d}k} \bigg).
\label{eq:eos_weight_definition}
\end{equation}
For the multifractional dispersion relation, this weight becomes
\begin{equation}
\mathcal{W}(k;E_{*}) = \frac{1}{3} \bigg(
\frac{ 1+5\sqrt{k/E_{*}}}{1+4\sqrt{k/E_{*}} } \bigg).
\label{eq:eos_weight_multifractional}
\end{equation}
{Since $W(k;E_{*})$ depends on the momentum, the equation of state is no longer characterized by a constant proportionality factor between the pressure and the energy density, as in the standard massless relativistic gas. Instead, different momentum modes contribute with different effective weights.}

We can also define the equation of state parameter which reads
\begin{equation}
w_{\sigma}(T,z) = \frac{P_{\sigma}}{u_{\sigma}}.
\label{eq:eos_parameter_definition}
\end{equation}
Using Eq.~\eqref{eq:pressure_weighted_form}, this can be written as
\begin{equation}
w_{\sigma}(T,z) = \frac{ \displaystyle \int_{0}^{\infty} k^{2}\omega(k)
\mathcal{W}(k;E_{*}) f_{\sigma}(k)
\,\mathrm{d}k}{ \displaystyle \int_{0}^{\infty}
k^{2}\omega(k) f_{\sigma}(k) \,\mathrm{d}k}.
\label{eq:eos_parameter_integral}
\end{equation}
In other words, $w_{\sigma}$ accounts for the thermal average of $\mathcal{W}(k;E_{*})$ weighted by the energy distribution of the gas.

Equation~\eqref{eq:eos_weight_multifractional} can also be written as
\begin{equation}
\mathcal{W}(k;E_{*}) = \frac{1}{3} +
\frac{1}{3} \bigg( \frac{\sqrt{k/E_{*}}}{1+4\sqrt{k/E_{*}}} \bigg).
\label{eq:eos_weight_correction}
\end{equation}
Consequently, for positive occupation numbers,
\begin{equation}
\frac{1}{3} \leq w_{\sigma}(T,z)
\leq \frac{5}{12}.
\label{eq:eos_bounds}
\end{equation}
The lower bound corresponds to the infrared relativistic regime, while the upper bound corresponds to the ultraviolet multifractional regime. In this first regime, the equation of state weight admits the expansion
\begin{equation}
\mathcal{W}(k;E_{*}) = \frac{1}{3}
+ \frac{1}{3}\sqrt{\frac{k}{E_{*}}} +
\mathcal{O} \left( \frac{k}{E_{*}} \right).
\label{eq:eos_weight_ir_expansion}
\end{equation}
Using the infrared expressions for $P_{\sigma}$ and $u_{\sigma}$, one finds
\begin{equation}
w_{\sigma}^{\mathrm{IR}}(T,z) \simeq
\frac{1}{3} \left[ 1 + \frac{35 \sqrt{\pi}}{32}
\sqrt{\frac{T}{E_{*}}} \frac{
\mathcal{L}_{9/2}^{(\sigma)}(z)}{ \mathcal{L}_{4}^{(\sigma)}(z)}\right]
+ \mathcal{O} \left( \frac{T}{E_{*}} \right).
\label{eq:w_ir_sigma}
\end{equation}
For radiation--like systems, we consider $z=1$. In this case, the leading multifractional correction shifts the relativistic value $w=1/3$ upward.

In the ultraviolet regime, the dispersion relation behaves as $\omega(k)\propto k^{5/4}$. More generally, for a power--law spectrum $\omega(k)\propto k^{\gamma}$, we have
$k\frac{\mathrm{d}\omega}{\mathrm{d}k} = \gamma \,\omega(k)$,
and therefore $P_{\sigma} = \frac{\gamma}{3}u_{\sigma}$.
For the present ultraviolet spectrum, $\gamma=5/4$, which gives $w_{\sigma}^{\mathrm{UV}} = \frac{5}{12}$.
This limiting value is independent of the quantum statistics; the statistics only affects how the system approaches this limit through the distribution function.

The same result follows directly from the ultraviolet thermodynamic functions.
Indeed, $P_{\sigma}^{\mathrm{UV}} \simeq
\mathcal{A} E_{*}^{3/5} T^{17/5}
\mathcal{L}_{17/5}^{(\sigma)}(z)$,
whereas $u_{\sigma}^{\mathrm{UV}} \simeq
\frac{12}{5} \mathcal{A} E_{*}^{3/5} T^{17/5}
\mathcal{L}_{17/5}^{(\sigma)}(z)$.
In this manner,
\begin{equation}
\frac{ P_{\sigma}^{\mathrm{UV}} }
{ u_{\sigma}^{\mathrm{UV}} } = \frac{5}{12}.
\label{eq:uv_pressure_density_ratio}
\end{equation}

It is also useful to introduce the adiabatic sound--speed parameter
\begin{equation}
c_{s,\sigma}^{2} = \left( \frac{\partial P_{\sigma}}{\partial u_{\sigma}} \right)_{z}.
\label{eq:sound_speed_definition}
\end{equation}
For the radiation--like sector, where the fugacity is fixed, the infrared expansion gives
\begin{equation}
c_{s,\sigma}^{2,\mathrm{IR}} \simeq
\frac{1}{3} \left[ 1 + \frac{315 \sqrt{\pi}}{256}
\sqrt{\frac{T}{E_{*}}} \frac{ \mathcal{L}_{9/2}^{(\sigma)}(z)}{ \mathcal{L}_{4}^{(\sigma)}(z) } \right] +
\mathcal{O} \left( \frac{T}{E_{*}} \right),
\label{eq:sound_speed_ir}
\end{equation}
while in the ultraviolet regime we get
\begin{equation}
c_{s,\sigma}^{2,\mathrm{UV}} = \frac{5}{12}.
\label{eq:sound_speed_uv}
\end{equation}
Here, notice that the multifractional correction makes the gas slightly stiffer than an ordinary relativistic gas, but the ultraviolet value remains below unity.


\section{Deformed Stefan--Boltzmann law }
\label{sec:deformed_stefan_boltzmann_law}

The Stefan--Boltzmann law is naturally associated with the radiation--like sector, for which the chemical potential vanishes. In the present notation this corresponds to $z=1$, as we have pointed out several times throughout the paper. The energy density is then obtained from
\begin{equation}
u_{\sigma}(T) = \frac{g}{2\pi^{2}}
\int_{0}^{\infty} \frac{ k^{2}\omega(k) }{ e^{\beta\omega(k)}+\sigma } \,\mathrm{d}k .
\label{eq:rho_radiation_sigma}
\end{equation}
For the standard massless gas, this expression gives $u\propto T^{4}$. In the multifractional case, the same integral receives corrections controlled by the ratio $T/E_{*}$.

It is useful to define an effective Stefan--Boltzmann coefficient
$\mathfrak{a}_{\sigma}(T/E_{*})$ through
\begin{equation}
u_{\sigma}(T) = \mathfrak{a}_{\sigma}\left(\frac{T}{E_{*}}\right)T^{4}.
\label{eq:effective_stefan_boltzmann_definition}
\end{equation}
Unlike the usual relativistic case, $\mathfrak{a}_{\sigma}$ is no longer a constant. Using the dimensionless integration variable introduced before, we obtain
\begin{equation}
\mathfrak{a}_{\sigma}\left(\frac{T}{E_{*}}\right) = \frac{g}{2\pi^{2}} \int_{0}^{\infty} \frac{ \xi^{3} \left( 1+4\sqrt{\xi T/E_{*}} \right)^{1/2} }
{ \exp\left[ \xi \left( 1+4\sqrt{\xi T/E_{*}} \right)^{1/2} \right] +\sigma} \,\mathrm{d}\xi .
\label{eq:effective_stefan_boltzmann_integral}
\end{equation}
This expression contains the full deformation of the thermal radiation law.

In the infrared regime, $T/E_{*}\ll1$, the energy density follows from Eq.~\eqref{eq:energy_density_ir_sigma} with $z=1$:
\begin{equation}
u_{\sigma}^{\mathrm{IR}}(T) \simeq
\frac{3gT^{4}}{\pi^{2}} \mathcal{L}_{4}^{(\sigma)}(1) - \frac{735g}{32\pi^{3/2}}
\frac{T^{9/2}}{E_{*}^{1/2}} \mathcal{L}_{9/2}^{(\sigma)}(1) + \mathcal{O} \left( \frac{T^{5}}{E_{*}} \right).
\label{eq:rho_ir_stefan_sigma}
\end{equation}
Equivalently,
\begin{equation}
u_{\sigma}^{\mathrm{IR}}(T) \simeq \frac{3gT^{4}}{\pi^{2}} \mathcal{L}_{4}^{(\sigma)}(1) \left[ 1 - \frac{245\sqrt{\pi}}{32} \sqrt{\frac{T}{E_{*}}}
\frac{ \mathcal{L}_{9/2}^{(\sigma)}(1)}{\mathcal{L}_{4}^{(\sigma)}(1)} + \mathcal{O}
\left( \frac{T}{E_{*}} \right)\right]
.
\label{eq:rho_ir_stefan_factorized}
\end{equation}
Then, the leading multifractional contribution lowers the standard Stefan--Boltzmann energy density for positive $E_{*}$.

For the bosonic sector, $\sigma=-1$, we have $\mathcal{L}_{s}^{(-)}(1) = \zeta(s)$,
where $\zeta(s)$ is the Riemann zeta function so that
\begin{equation}
u_{B}^{\mathrm{IR}}(T) \simeq
\frac{g\pi^{2}}{30}T^{4} -
\frac{735g}{32\pi^{3/2}} \zeta\left(\frac{9}{2}\right) \frac{T^{9/2}}{E_{*}^{1/2}}
+ \mathcal{O} \left( \frac{T^{5}}{E_{*}} \right).
\label{eq:rho_ir_boson_stefan}
\end{equation}
The first term is the usual bosonic Stefan--Boltzmann law, while the second term is the leading multifractional correction.

On the other hand, for the fermionic sector, $\sigma=+1$, we have $\mathcal{L}_{s}^{(+)}(1) =
\left( 1-2^{1-s} \right)\zeta(s)$.
Then,
\begin{equation}
u_{F}^{\mathrm{IR}}(T) \simeq
\frac{7g\pi^{2}}{240}T^{4} -
\frac{735g}{32\pi^{3/2}} \left( 1-2^{-7/2} \right)
\zeta\left(\frac{9}{2}\right)
\frac{T^{9/2}}{E_{*}^{1/2}} +
\mathcal{O} \left( \frac{T^{5}}{E_{*}} \right).
\label{eq:rho_ir_fermion_stefan}
\end{equation}
The usual factor $7/8$ relating fermionic and bosonic radiation is recovered in the leading $T^{4}$ term.

The ultraviolet regime displays a different power law. From Eq.~\eqref{eq:energy_density_uv_sigma} with $z=1$, we are able to write $ u_{\sigma}^{\mathrm{UV}}(T) \simeq
\frac{12}{5} \mathcal{A}
E_{*}^{3/5} T^{17/5} \mathcal{L}_{17/5}^{(\sigma)}(1)$.
Equivalently, using the explicit form of $\mathcal{A}$,
\begin{equation}
u_{\sigma}^{\mathrm{UV}}(T) \simeq
\frac{ 12\,\times 2^{-7/5}g}{
25\pi^{2}} \Gamma\left(\frac{12}{5}\right)
E_{*}^{3/5} T^{17/5} \mathcal{L}_{17/5}^{(\sigma)}(1).
\label{eq:rho_uv_stefan_sigma}
\end{equation}
Therefore, the ordinary scaling $u(T)\propto T^{4}$
is replaced, in the ultraviolet sector, by $u_{\sigma}^{\mathrm{UV}}(T)
\propto E_{*}^{3/5}T^{17/5}$.
The exponent $17/5$ reflects the ultraviolet density of states behavior induced
by the multifractional dispersion relation.

For bosons, the ultraviolet law becomes
\begin{equation}
u_{B}^{\mathrm{UV}}(T) \simeq
\frac{ 12\,2^{-7/5}g}{25\pi^{2}}
\Gamma\left(\frac{12}{5}\right) \zeta\left(\frac{17}{5}\right)
E_{*}^{3/5} T^{17/5}.
\label{eq:rho_uv_boson_stefan}
\end{equation}
For fermions, we write
\begin{equation}
u_{F}^{\mathrm{UV}}(T) \simeq
\frac{ 12\,2^{-7/5}g}{25\pi^{2}}
\Gamma\left(\frac{12}{5}\right)
\left( 1-2^{-12/5} \right)
\zeta\left(\frac{17}{5}\right)
E_{*}^{3/5} T^{17/5}.
\label{eq:rho_uv_fermion_stefan}
\end{equation}

In terms of the effective coefficient introduced in
Eq.~\eqref{eq:effective_stefan_boltzmann_definition}, the ultraviolet regime
corresponds to
\begin{equation}
\mathfrak{a}_{\sigma}^{\mathrm{UV}}
\left( \frac{T}{E_{*}} \right) \simeq
\frac{ 12\,2^{-7/5}g}{25\pi^{2}}
\Gamma\left(\frac{12}{5}\right) \mathcal{L}_{17/5}^{(\sigma)}(1)
\left( \frac{E_{*}}{T} \right)^{3/5}.
\label{eq:effective_stefan_uv}
\end{equation}
The energy density still increases with temperature, but more slowly than in the ordinary relativistic gas at sufficiently high temperature. On the other hand, the corresponding pressure follows from the ultraviolet equation of state,
\begin{equation}
P_{\sigma}^{\mathrm{UV}} = \frac{5}{12} u_{\sigma}^{\mathrm{UV}},
\label{eq:pressure_uv_stefan}
\end{equation}
rather than the usual radiation relation $P=u/3$. 

{It is useful to compare these results with the earlier black--body
analysis of the multifractional theory with $q$--derivatives presented
in Ref.~\cite{Calcagni:2016CMB}. In that work, the thermal ensemble is
formulated in terms of the geometric energy and its conjugate geometric
inverse temperature, leading to a radiation density proportional to
$[\beta_q(T)]^{-4}$. By contrast, the present analysis treats
$\omega(k)$ as an effective physical spectrum within the standard
grand--canonical ensemble. Both prescriptions recover the ordinary
$T^4$ behavior and predict a negative leading correction in the
infrared. However, they need not yield the same coefficient or the same
ultraviolet scaling. In particular, for $\alpha_0=1/2$, the geometric
construction of Ref.~\cite{Calcagni:2016CMB} gives
\[
\rho_q^{\rm UV}(T)\propto E_*^2T^2,
\]
whereas the effective-spectrum result obtained in Eq.~(\ref{eq:rho_uv_stefan_sigma}) scales as
$E_*^{3/5}T^{17/5}$. This difference follows from the distinct
thermodynamic prescriptions and not from the spatial momentum measure.
Accordingly, the ultraviolet results obtained here should be understood
as properties of the effective dispersion relation considered in this
work, rather than as a rederivation of the complete geometric
thermodynamics of the $T_q$ theory.}


\section{Thermal stability and fluctuations }
\label{sec:thermal_stability_fluctuations}

We now examine the response functions and fluctuation sector of the multifractional gas. The particle number fluctuation is defined by
\begin{equation}
\left\langle
(\Delta N)^{2}
\right\rangle_{\sigma} = \left\langle
N^{2} \right\rangle_{\sigma} - \left\langle
N \right\rangle_{\sigma}^{2}.
\label{eq:number_fluctuation_definition}
\end{equation}
Using the grand--canonical distribution, we get
\begin{equation}
\frac{ \left\langle (\Delta N)^{2} \right\rangle_{\sigma} }{V} =\frac{g}{2\pi^{2}}
\int_{0}^{\infty} k^{2} f_{\sigma}(k)
\left[ 1-\sigma f_{\sigma}(k) \right] \mathrm{d}k .
\label{eq:number_fluctuation_density}
\end{equation}
Equivalently,
\begin{equation}
\frac{ \left\langle (\Delta N)^{2} \right\rangle_{\sigma} }{V} = T \left( \frac{\partial n_{\sigma}}{\partial \mu} \right)_{T} = z \left( \frac{\partial n_{\sigma}}{\partial z} \right)_{T}.
\label{eq:number_fluctuation_susceptibility}
\end{equation}
The quantity $\chi_{T}^{(\sigma)} = \left( \frac{\partial n_{\sigma}}{\partial \mu} \right)_{T}$ is the isothermal number susceptibility. Furthermore, the local stability with respect to particle number fluctuations requires $\chi_{T}^{(\sigma)}>0$.

The energy fluctuation is defined by
\begin{equation}
\left\langle (\Delta E)^{2} \right\rangle_{\sigma}
= \left\langle
E^{2} \right\rangle_{\sigma} - U_{\sigma}^{2},
\label{eq:energy_fluctuation_definition}
\end{equation}
where $E$ denotes the fluctuating total energy of a microscopic configuration. For an ideal quantum gas, the independent--mode structure gives
\begin{equation}
\frac{ \left\langle
(\Delta E)^{2} \right\rangle_{\sigma}}{V}
= \frac{g}{2\pi^{2}}
\int_{0}^{\infty} k^{2}
\omega^{2}(k) f_{\sigma}(k)
\left[ 1-\sigma f_{\sigma}(k) \right] \mathrm{d}k .
\label{eq:energy_fluctuation_density}
\end{equation}
At fixed fugacity, this fluctuation is related to the heat capacity density by
\begin{equation}
\frac{ \left\langle (\Delta E)^{2} \right\rangle_{\sigma}}{V}
= T^{2} c_{V,z}^{(\sigma)},
\label{eq:energy_fluctuation_cvz}
\end{equation}
where $c_{V,z}^{(\sigma)} = \left( \frac{\partial u_{\sigma}}{\partial T} \right)_{z}$. The positivity of the energy fluctuation implies $c_{V,z}^{(\sigma)}>0$. For the radiation--like sector, $z=1$, and the fixed fugacity heat capacity is the natural response function.

The mixed fluctuation between energy and particle number is
\begin{equation}
\left\langle \Delta E\,\Delta N \right\rangle_{\sigma}
= \left\langle
EN \right\rangle_{\sigma} - U_{\sigma}
\left\langle N \right\rangle_{\sigma}.
\label{eq:energy_number_covariance_definition}
\end{equation}
It is given by
\begin{equation}
\frac{ \left\langle \Delta E\,\Delta N
\right\rangle_{\sigma} }{V}
= \frac{g}{2\pi^{2}} \int_{0}^{\infty}
k^{2} \omega(k) f_{\sigma}(k) \left[ 1-\sigma f_{\sigma}(k) \right] \mathrm{d}k ,
\label{eq:energy_number_covariance}
\end{equation}
or equivalently,
\begin{equation}
\frac{ \left\langle \Delta E\,\Delta N
\right\rangle_{\sigma}}{V} =
T \left( \frac{\partial u_{\sigma}}{\partial \mu}
\right)_{T}
= z \left( \frac{\partial u_{\sigma}}{\partial z}\right)_{T}.
\label{eq:energy_number_covariance_derivative}
\end{equation}

The fluctuation matrix per unit volume can be written as
\begin{equation}
\mathcal{C}_{\sigma} = \frac{1}{V}
\begin{pmatrix}
\left\langle(\Delta E)^{2}\right\rangle_{\sigma}
&
\left\langle\Delta E\,\Delta N\right\rangle_{\sigma}
\\[2mm]
\left\langle\Delta E\,\Delta N\right\rangle_{\sigma}
&
\left\langle(\Delta N)^{2}\right\rangle_{\sigma}
\end{pmatrix}.
\label{eq:fluctuation_matrix}
\end{equation}
Thermodynamic stability requires this matrix to be positive semidefinite so that we can write
$\left\langle(\Delta E)^{2}\right\rangle_{\sigma}\geq0$, $\left\langle(\Delta N)^{2}\right\rangle_{\sigma}\geq0$,
together with $\left\langle(\Delta E)^{2}\right\rangle_{\sigma}
\left\langle(\Delta N)^{2}\right\rangle_{\sigma}
- \left\langle\Delta E\,\Delta N\right\rangle_{\sigma}^{2} \geq0$. In other words, for the ideal gas considered here, these conditions follow from the positivity
of the factor $f_{\sigma}(k)[1-\sigma f_{\sigma}(k)]$ in the allowed thermodynamic domain.

The distinction between fermionic and bosonic fluctuations is explicit in Eq.~\eqref{eq:number_fluctuation_density}. For fermions, $f_{+}(k)
\left[ 1-f_{+}(k) \right]$ is suppressed by the Pauli exclusion principle. For bosons,
$f_{-}(k) \left[ 1+f_{-}(k) \right]$
contains the Bose enhancement factor. Close to the bosonic condensation regime, this enhancement becomes important and must be treated separately from the contribution of the continuum modes.

In the infrared regime, the particle number fluctuation follows from the low--temperature expansion of $n_{\sigma}$. Using
$z \frac{\partial}{\partial z} \mathcal{L}_{s}^{(\sigma)}(z) = \mathcal{L}_{s-1}^{(\sigma)}(z)$,
we obtain
\begin{equation}
\frac{ \left\langle
(\Delta N)^{2} \right\rangle_{\sigma}^{\mathrm{IR}}}{V}
\simeq \frac{gT^{3}}{\pi^{2}} \mathcal{L}_{2}^{(\sigma)}(z)
- \frac{105g}{16\pi^{3/2}} \frac{T^{7/2}}{E_{*}^{1/2}} \mathcal{L}_{5/2}^{(\sigma)}(z)
+ \mathcal{O} \left( \frac{T^{4}}{E_{*}} \right).
\label{eq:number_fluctuation_ir}
\end{equation}
The corresponding mixed fluctuation is
\begin{equation}
\frac{ \left\langle
\Delta E\,\Delta N \right\rangle_{\sigma}^{\mathrm{IR}} }{V}
\simeq \frac{3gT^{4}}{\pi^{2}}
\mathcal{L}_{3}^{(\sigma)}(z)
- \frac{735g}{32\pi^{3/2}}
\frac{T^{9/2}}{E_{*}^{1/2}}
\mathcal{L}_{7/2}^{(\sigma)}(z)
+ \mathcal{O}
\left( \frac{T^{5}}{E_{*}} \right).
\label{eq:energy_number_covariance_ir}
\end{equation}
At fixed fugacity, the energy fluctuation becomes
\begin{equation}
\frac{ \left\langle (\Delta E)^{2}
\right\rangle_{\sigma}^{\mathrm{IR}} }{V}
\simeq \frac{12gT^{5}}{\pi^{2}}
\mathcal{L}_{4}^{(\sigma)}(z)
- \frac{6615g}{64\pi^{3/2}} \frac{T^{11/2}}{E_{*}^{1/2}} \mathcal{L}_{9/2}^{(\sigma)}(z)
+ \mathcal{O}
\left( \frac{T^{6}}{E_{*}} \right).
\label{eq:energy_fluctuation_ir}
\end{equation}
Therefore, the leading multifractional correction reduces the infrared fluctuation amplitudes for positive $E_{*}$, consistently with the reduction found in the energy density.

In the ultraviolet regime, the particle number  fluctuation is
\begin{equation}
\frac{ \left\langle
(\Delta N)^{2} \right\rangle_{\sigma}^{\mathrm{UV}}}{V}
\simeq \mathcal{A} E_{*}^{3/5} T^{12/5} \mathcal{L}_{7/5}^{(\sigma)}(z).
\label{eq:number_fluctuation_uv}
\end{equation}
The energy--particle covariance is
\begin{equation}
\frac{
\left\langle \Delta E\,\Delta N
\right\rangle_{\sigma}^{\mathrm{UV}} }{V}
\simeq \frac{12}{5} \mathcal{A} E_{*}^{3/5} T^{17/5} \mathcal{L}_{12/5}^{(\sigma)}(z),
\label{eq:energy_number_covariance_uv}
\end{equation}
and the energy fluctuation is
\begin{equation}
\frac{ \left\langle (\Delta E)^{2}
\right\rangle_{\sigma}^{\mathrm{UV}}}{V}
\simeq \frac{204}{25} \mathcal{A} E_{*}^{3/5}
T^{22/5} \mathcal{L}_{17/5}^{(\sigma)}(z).
\label{eq:energy_fluctuation_uv}
\end{equation}
In the high--temperature multifractional regime, the fluctuation sector inherits the same nonstandard thermal powers that characterize the modified Stefan--Boltzmann law.

For $z=1$, the ultraviolet heat capacity density is
\begin{equation}
c_{V}^{(\sigma),\mathrm{UV}} \simeq
\frac{204}{25} \mathcal{A}
E_{*}^{3/5} T^{12/5} \mathcal{L}_{17/5}^{(\sigma)}(1).
\label{eq:cv_uv_fluctuation_section}
\end{equation}
Since $\mathcal{A}>0$ and $\mathcal{L}_{17/5}^{(\sigma)}(1)>0$ for both admissible bosonic and fermionic radiation sectors, the ultraviolet gas is locally thermally stable. This stability is compatible with the equation of state bound $1/3\leq w_{\sigma}\leq5/12$, which keeps the sound-speed parameter below unity in the limiting regimes considered above.


\section{Bosonic sector and Bose--Einstein condensation }
\label{sec:bosonic_sector_bec}

We now specialize the grand--canonical formulation to the bosonic sector, corresponding to $\sigma=-1$. Since the spectrum is gapless, the lowest energy is $\omega_{0}=0$, and the
chemical potential must satisfy $\mu\leq0$,  $0<z\leq1$. The endpoint $z=1$ corresponds to the saturation of the excited states. In addition, for bosons with conserved particle number, the total number of particles must include the occupation of the zero momentum mode. In this manner, we write
\begin{equation}
N = N_{0} + N_{\mathrm{ex}},
\label{eq:total_number_bec}
\end{equation}
where $N_{0}$ is the ground state occupation number and $N_{\mathrm{ex}}$ is the number of particles in the continuum of excited states. For a nondegenerate ground state with $\omega_{0}=0$, we obtain $N_{0} = z/(1-z)$. The excited contribution is
\begin{equation}
N_{\mathrm{ex}} = \frac{gV}{2\pi^{2}}
\int_{0}^{\infty} \frac{k^{2}}
{z^{-1}e^{\beta\omega(k)}-1} \,\mathrm{d}k .
\label{eq:excited_number_bec}
\end{equation}
Equivalently, the excited number density is
$n_{\mathrm{ex}} = N_{\mathrm{ex}}/V$.

The Bose--Einstein condensation temperature is obtained from the saturation condition
$n = n_{\mathrm{ex}}(T_{c},z=1)$,
where $T_{c}$ denotes the critical temperature. Notice that this equation determines $T_{c}$ implicitly for the multifractional spectrum.
In the infrared region, $\omega(k)\simeq k$, and the integrand behaves as
\begin{equation}
\frac{k^{2}}
{e^{\omega(k)/T_{c}}-1} \simeq T_{c}k, \qquad k\rightarrow0.
\label{eq:bec_ir_convergence}
\end{equation}
Therefore, the low momentum sector does not produce an infrared divergence. The ultraviolet sector is exponentially suppressed by the Bose--Einstein factor.

Using the infrared expansion of the number density, the critical density is
\begin{equation}
n_{c}^{\mathrm{IR}}(T) \simeq \frac{gT^{3}}{\pi^{2}} \zeta(3) - \frac{105g}{16\pi^{3/2}}
\zeta\left(\frac{7}{2}\right) \frac{T^{7/2}}{E_{*}^{1/2}} + \mathcal{O} \left( \frac{T^{4}}{E_{*}} \right).
\label{eq:critical_density_ir_bec}
\end{equation}
The first term is the standard relativistic contribution, while the second term is the leading multifractional correction.

Let $T_{0}$ be the critical temperature obtained in the undeformed limit, defined by
\begin{equation}
T_{0} = \left[ \frac{\pi^{2}n} {g\zeta(3)} \right]^{1/3}.
\label{eq:T0_bec}
\end{equation}
Solving Eq.~\eqref{eq:critical_density_ir_bec} perturbatively for fixed $n$, we are able to write
\begin{equation}
T_{c} \simeq T_{0} \left[ 1 + \frac{35\sqrt{\pi}}{16} \frac{ \zeta\left(\frac{7}{2}\right)}{
\zeta(3) } \sqrt{\frac{T_{0}}{E_{*}}} \right].
\label{eq:Tc_ir_bec}
\end{equation}
Remarkably, for positive $E_{*}$, the leading multifractional correction increases the critical temperature. This occurs because the modified density of states reduces the excited state population at fixed temperature, so saturation is reached at a higher value of $T$ for fixed total density.

Below the critical temperature, the fugacity remains pinned at $z=1$, and the excess particles occupy the zero--momentum state. The condensate fraction is
\begin{equation}
\frac{N_{0}}{N} = 1 - \frac{n_{\mathrm{ex}}(T,z=1)}{n}.
\label{eq:condensate_fraction_definition}
\end{equation}
Using the infrared critical density, this gives
\begin{equation}
\frac{N_{0}}{N} \simeq 1 - \left( \frac{T}{T_{c}}
\right)^{3} \frac{ 1 - \eta \sqrt{T/E_{*}}}{1
- \eta \sqrt{T_{c}/E_{*}}},
\label{eq:condensate_fraction_ir}
\end{equation}
where the dimensionless coefficient $\eta$ is defined by
\begin{equation}
\eta = \frac{105\sqrt{\pi}}{16} \frac{ \zeta\left(\frac{7}{2}\right) }{ \zeta(3) }.
\label{eq:eta_bec_definition}
\end{equation}
In the limit $E_{*}\rightarrow\infty$, Eq.~\eqref{eq:condensate_fraction_ir} reduces to the usual relativistic result
\begin{equation}
\frac{N_{0}}{N} = 1 - \left( \frac{T}{T_{c}} \right)^{3}.
\label{eq:standard_condensate_fraction}
\end{equation}

The thermodynamic quantities below $T_{c}$ are determined by the excited states, since the zero mode has $\omega_{0}=0$ and does not contribute to the energy or pressure. Therefore,
\begin{equation}
u_{B}(T<T_{c}) = \frac{g}{2\pi^{2}} \int_{0}^{\infty} \frac{k^{2}\omega(k)} {e^{\beta\omega(k)}-1} \,\mathrm{d}k,
\label{eq:rho_below_Tc_bec}
\end{equation}
and
\begin{equation}
P_{B}(T<T_{c}) = \frac{T}{V} \ln\mathcal{Z}_{B}^{\mathrm{ex}},
\label{eq:pressure_below_Tc_bec}
\end{equation}
where $\mathcal{Z}_{B}^{\mathrm{ex}}$ denotes the bosonic partition function restricted to the excited modes. In the infrared regime, these quantities are given by
\begin{equation}
u_{B}^{\mathrm{IR}}(T<T_{c}) \simeq
\frac{g\pi^{2}}{30}T^{4} - \frac{735g}{32\pi^{3/2}}
\zeta\left(\frac{9}{2}\right)
\frac{T^{9/2}}{E_{*}^{1/2}} + \mathcal{O}
\left( \frac{T^{5}}{E_{*}} \right),
\label{eq:rho_bec_ir}
\end{equation}
and
\begin{equation}
P_{B}^{\mathrm{IR}}(T<T_{c}) \simeq
\frac{g\pi^{2}}{90}T^{4} - \frac{105g}{16\pi^{3/2}}
\zeta\left(\frac{9}{2}\right) \frac{T^{9/2}}{E_{*}^{1/2}} + \mathcal{O}
\left( \frac{T^{5}}{E_{*}} \right).
\label{eq:pressure_bec_ir}
\end{equation}
The condensate changes the particle number balance but does not add a direct zero mode contribution to these two quantities.

In the ultraviolet regime, the critical density becomes
\begin{equation}
n_{c}^{\mathrm{UV}}(T) \simeq \mathcal{A}
E_{*}^{3/5} T^{12/5} \zeta\left(\frac{12}{5}\right).
\label{eq:critical_density_uv_bec}
\end{equation}
Consequently, the ultraviolet critical temperature is
\begin{equation}
T_{c}^{\mathrm{UV}} \simeq \left[ \frac{ n}{
\mathcal{A} E_{*}^{3/5} \zeta\left(\frac{12}{5}\right)} \right]^{5/12}.
\label{eq:Tc_uv_bec}
\end{equation}
The associated condensate fraction in this limiting regime is
\begin{equation}
\frac{N_{0}}{N} \simeq 1 - \left( \frac{T}{T_{c}^{\mathrm{UV}}} \right)^{12/5}.
\label{eq:condensate_fraction_uv}
\end{equation}
The exponent $12/5$ reflects the effective density of states dimension induced by the multifractional spectrum.

It is important to distinguish this conserved bosonic gas from a radiation--like gas. For photons or gravitons in thermal equilibrium, the particle number is not fixed, and one sets $\mu=0$ from the beginning. In that case, $z=1$ is not a saturation condition imposed by number conservation, and the standard Bose--Einstein condensation interpretation does not apply. The condensation analysis above is therefore appropriate for a conserved bosonic sector or for an effective quasiparticle system governed by the same multifractional dispersion relation.


\section{Fermionic sector and degenerate Fermi gas}
\label{sec:fermionic_sector_degenerate_gas}

We now consider the fermionic sector, obtained by setting $\sigma=+1$. In the degenerate regime, the temperature is much smaller than the chemical potential. In the zero--temperature limit, the distribution becomes $f_{F}(k) \longrightarrow \Theta(k_{F}-k)$, where $\Theta$ is the Heaviside step function and $k_{F}$ is the Fermi momentum. The number density is then
\begin{equation}
n = \frac{g}{2\pi^{2}} \int_{0}^{k_{F}} k^{2}\,\mathrm{d}k = \frac{g}{6\pi^{2}}k_{F}^{3}.
\label{eq:fermi_number_density}
\end{equation}
In this way, $k_{F} = \left( \frac{6\pi^{2}n}{g} \right)^{1/3}$.

The Fermi energy is defined by the value of the modified spectrum at the Fermi surface, $\varepsilon_{F} = \omega(k_{F})$. For the multifractional dispersion relation, this gives
\begin{equation}
\varepsilon_{F} = k_{F} \left( 1+4\sqrt{\frac{k_{F}}{E_{*}}} \right)^{1/2}.
\label{eq:fermi_energy_multifractional}
\end{equation}
In the infrared regime, $k_{F}/E_{*}\ll1$, we get
\begin{equation}
\varepsilon_{F} \simeq k_{F} + 2\frac{k_{F}^{3/2}}{E_{*}^{1/2}} - 2\frac{k_{F}^{2}}{E_{*}} +
\mathcal{O} \left( \frac{k_{F}^{5/2}}{E_{*}^{3/2}} \right).
\label{eq:fermi_energy_ir}
\end{equation}
In other words, the multifractional correction raises the Fermi energy with respect to the standard massless result.

At zero temperature, the energy density is
\begin{equation}
u_{F}^{(0)} = \frac{g}{2\pi^{2}} \int_{0}^{k_{F}} k^{2}\omega(k)\,\mathrm{d}k .
\label{eq:fermi_energy_density_zeroT}
\end{equation}
The corresponding pressure can be obtained from the kinetic expression
\begin{equation}
P_{F}^{(0)} = \frac{g}{6\pi^{2}} \int_{0}^{k_{F}}
k^{3} \frac{\mathrm{d}\omega}{\mathrm{d}k} \,\mathrm{d}k .
\label{eq:fermi_pressure_zeroT}
\end{equation}
Equivalently, using the zero--temperature thermodynamic relation, we may write
$P_{F}^{(0)} = n\varepsilon_{F} - u_{F}^{(0)}$.

Using the infrared expansion of the spectrum, Eq.~\eqref{eq:fermi_energy_density_zeroT} gives
\begin{equation}
u_{F}^{(0),\mathrm{IR}} \simeq \frac{g}{8\pi^{2}}k_{F}^{4} + \frac{2g}{9\pi^{2}}
\frac{k_{F}^{9/2}}{E_{*}^{1/2}} - \frac{g}{5\pi^{2}} \frac{k_{F}^{5}}{E_{*}}
+ \mathcal{O} \left( \frac{k_{F}^{11/2}}{E_{*}^{3/2}} \right).
\label{eq:fermi_energy_density_ir}
\end{equation}
Similarly, the pressure becomes
\begin{equation}
P_{F}^{(0),\mathrm{IR}} \simeq
\frac{g}{24\pi^{2}}k_{F}^{4} + \frac{g}{9\pi^{2}}
\frac{k_{F}^{9/2}}{E_{*}^{1/2}} - \frac{2g}{15\pi^{2}} \frac{k_{F}^{5}}{E_{*}}
+ \mathcal{O} \left( \frac{k_{F}^{11/2}}{E_{*}^{3/2}} \right).
\label{eq:fermi_pressure_ir}
\end{equation}
The leading terms satisfy the usual massless relation $P_{F}^{(0)}=u_{F}^{(0)}/3$, whereas the multifractional terms deform the degenerate equation of state.

In the ultraviolet regime, the spectrum behaves as $\omega(k)\propto k^{5/4}$. The Fermi energy becomes
\begin{equation}
\varepsilon_{F}^{\mathrm{UV}} \simeq 2E_{*}^{-1/4}k_{F}^{5/4}.
\label{eq:fermi_energy_uv}
\end{equation}
The energy density is then
\begin{equation}
u_{F}^{(0),\mathrm{UV}} \simeq
\frac{4g}{17\pi^{2}} E_{*}^{-1/4} k_{F}^{17/4},
\label{eq:fermi_energy_density_uv}
\end{equation}
while the pressure is
\begin{equation}
P_{F}^{(0),\mathrm{UV}} \simeq \frac{5g}{51\pi^{2}}
E_{*}^{-1/4} k_{F}^{17/4}.
\label{eq:fermi_pressure_uv}
\end{equation}
Consequently,
\begin{equation}
P_{F}^{(0),\mathrm{UV}} = \frac{5}{12} u_{F}^{(0),\mathrm{UV}}.
\label{eq:fermi_uv_eos}
\end{equation}
This is the same ultraviolet equation of state ratio obtained for the thermal gas, showing that the high--energy scaling is controlled by the power of the dispersion relation rather than by the quantum statistics.

The zero--temperature grand potential density is
\begin{equation}
\frac{\Phi_{F}^{(0)}}{V} = u_{F}^{(0)}
- \mu n,
\label{eq:fermi_grand_potential_zeroT}
\end{equation}
with $\mu=\varepsilon_{F}$ at $T=0$. Then,
\begin{equation}
\frac{\Phi_{F}^{(0)}}{V} = -P_{F}^{(0)}.
\label{eq:fermi_grand_potential_pressure_zeroT}
\end{equation}
Notice that this relation is useful for connecting the degenerate limit with the grand--canonical construction developed above.

The stiffness of the degenerate gas can be characterized by the zero--temperature sound--speed parameter $c_{s,F}^{2} = \left( \frac{\mathrm{d}P_{F}^{(0)}}{\mathrm{d}u_{F}^{(0)}} \right)$. Differentiating with respect to $k_{F}$ gives
\begin{equation}
c_{s,F}^{2} = \frac{1}{3} \frac{k_{F}}{\omega(k_{F})} \left. \frac{\mathrm{d}\omega}{\mathrm{d}k} \right|_{k=k_{F}} .
\label{eq:fermi_sound_speed_general}
\end{equation}
For the multifractional spectrum,
\begin{equation}
c_{s,F}^{2} = \frac{1}{3} \bigg( \frac{ 1+5\sqrt{k_{F}/E_{*}} }{ 1+4\sqrt{k_{F}/E_{*}}} \bigg).
\label{eq:fermi_sound_speed_multifractional}
\end{equation}
Therefore,
\begin{equation}
\frac{1}{3} \leq c_{s,F}^{2} \leq \frac{5}{12}.
\label{eq:fermi_sound_speed_bounds}
\end{equation}
The lower bound is recovered in the standard infrared limit, while the upper bound is approached in the ultraviolet multifractional regime. In addition, finite temperature corrections can be extracted through a Sommerfeld expansion. Let
\begin{equation}
\nu_{F} = \frac{1}{V}
\varrho(\varepsilon_{F}) = \frac{g}{2\pi^{2}}
\frac{k_{F}^{2}}{ v_{g}(k_{F})}
\label{eq:fermi_surface_dos}
\end{equation}
be the density of states per unit volume evaluated at the Fermi surface. At fixed number density and for $T\ll\varepsilon_{F}$, the leading thermal correction to the energy density is
\begin{equation}
u_{F}(T) \simeq u_{F}^{(0)} + \frac{\pi^{2}}{6}\nu_{F}T^{2}.
\label{eq:fermi_lowT_energy_density}
\end{equation}
The heat capacity density at fixed number density is therefore
\begin{equation}
c_{V,n}^{(F)} = \left( \frac{\partial u_{F}}{\partial T} \right)_{n} \simeq \frac{\pi^{2}}{3}\nu_{F}T.
\label{eq:fermi_lowT_heat_capacity}
\end{equation}
The linear dependence on $T$ is the usual signature of a degenerate Fermi gas, but the coefficient is deformed through the modified Fermi surface density of states.

Using the explicit group velocity, Eq.~\eqref{eq:fermi_surface_dos} becomes
\begin{equation}
\nu_{F} = \frac{g}{2\pi^{2}} k_{F}^{2}
\frac{ \left( 1+4\sqrt{k_{F}/E_{*}} \right)^{1/2}} { 1+5\sqrt{k_{F}/E_{*}}}.
\label{eq:fermi_surface_dos_multifractional}
\end{equation}
In the infrared regime,
\begin{equation}
\nu_{F}^{\mathrm{IR}} \simeq
\frac{g}{2\pi^{2}}k_{F}^{2} \left[ 1 - 3\sqrt{\frac{k_{F}}{E_{*}}} +
\mathcal{O} \left( \frac{k_{F}}{E_{*}} \right) \right],
\label{eq:fermi_surface_dos_ir}
\end{equation}
whereas in the ultraviolet regime,
\begin{equation}
\nu_{F}^{\mathrm{UV}} \simeq \frac{g}{5\pi^{2}}
E_{*}^{1/4} k_{F}^{7/4}.
\label{eq:fermi_surface_dos_uv}
\end{equation}
Remarkably, what we have got is that the multifractional correction modifies not only the zero--temperature equation of state, but also the low--temperature response of the degenerate fermionic system.


{

\section{Phenomenological bounds and measurable consequences}
\label{sec:phenomenological-bounds}

The results derived above have three complementary expansion parameters. Thermal observables are controlled by $T/E_*$, spectral observables by $\omega/E_*$, and the degenerate fermionic sector by $k_F/E_*$. In this section, we translate the corresponding corrections into measurable quantities and compare the characteristic scale $E_*$ with laboratory and astrophysical constraints. A distinction is essential from the outset. If the dispersion relation is the universal one--particle spectrum of the multifractional theory with $q$--derivatives, the same value of $E_*$ can be constrained by photons, gravitons, and atomic physics. If, instead, it is interpreted as the effective spectrum of a particular quasiparticle medium, bounds obtained in other sectors do not apply. In the latter case, $E_*$ is a medium--dependent matching scale that must be extracted from the microscopic band or collective excitation. We consider these two interpretations separately and indicate the physical consequence of each thermodynamic result.


\subsection{Density of states, response functions, and infrared observables}
\label{subsec:infrared-observables}

For the timelike branch studied here, the dispersion relation and its group velocity are
\begin{equation}
\nonumber
\omega^2=k^2+4E_*^{-1/2}k^{5/2}, \qquad v_{\rm g}=\frac{\mathrm{d}\omega}{\mathrm{d}k} =\frac{1+5\sqrt{k/E_*}}{\sqrt{1+4\sqrt{k/E_*}}} =1+3\sqrt{\frac{k}{E_*}}+\mathcal{O}\!\left(\frac{k}{E_*}\right).
\label{eq:group-velocity-bound-section}
\end{equation}
The density of one--particle states per unit volume is therefore
\begin{equation}
\frac{\varrho(\omega)}{V}
=\frac{g}{2\pi^2}\frac{k^2(\omega)}{v_{\rm g}[k(\omega)]}
=\frac{g\omega^2}{2\pi^2}
\left[1-7\sqrt{\frac{\omega}{E_*}}
+\mathcal{O}\!\left(\frac{\omega}{E_*}\right)\right].
\label{eq:dos-observable-ir}
\end{equation}
This expression gives a direct physical interpretation of the modified density of states. At a fixed energy, the deformation reduces the number of available modes. This reduction is inherited by the spectral intensity, the thermodynamic densities, and their fluctuations. In the opposite limit, $\varrho(\omega)\propto E_*^{3/5}\omega^{7/5}$, corresponding to $d_{\rm eff}=12/5$. The ultraviolet gas therefore contains fewer thermally accessible states than a standard relativistic gas at the same energy.

For the bosonic radiation sector, the leading infrared corrections to all the principal equilibrium quantities can be written as
\begin{equation}
\frac{Q_B-Q_B^{(0)}}{Q_B^{(0)}} =-C_Q\sqrt{\frac{T}{E_*}} +\mathcal{O}\!\left(\frac{T}{E_*}\right), \qquad Q_B\in\{P_B,n_B,u_B,s_B,c_{V,B}\}.
\label{eq:relative-thermal-shifts}
\end{equation}
The corresponding coefficients are
\begin{align}
C_P&=\frac{105\sqrt{\pi}\,\zeta(9/2)}{16\zeta(4)}\simeq11.335, & C_n&=\frac{105\sqrt{\pi}\,\zeta(7/2)}{16\zeta(3)}\simeq10.903, \nonumber\\ C_u&=\frac{245\sqrt{\pi}\,\zeta(9/2)}{32\zeta(4)}\simeq13.224, & C_s&=\frac{945\sqrt{\pi}\,\zeta(9/2)}{128\zeta(4)}\simeq12.752, \nonumber\\ C_V&=\frac{2205\sqrt{\pi}\,\zeta(9/2)}{256\zeta(4)}\simeq14.877.
\label{eq:thermal-shift-coefficients}
\end{align}
In this manner, the deformation does not simply change the Stefan--Boltzmann energy density. It simultaneously lowers the pressure, particle number density, entropy density, and heat capacity. Their relative shifts have fixed ratios. For instance, $\Delta c_{V,B}/c_{V,B}^{(0)}=(9/8)\Delta u_B/u_B^{(0)}$ at leading order. This correlated pattern is more characteristic than a change in a single normalization and could address a consistency test for an effective quasiparticle realization.

The equation of state and the adiabatic sound--speed parameter shift in the opposite direction:
\begin{align}
w &=\frac{1}{3}\left[1+C_w\sqrt{\frac{T}{E_*}} +\mathcal{O}\!\left(\frac{T}{E_*}\right)\right], & C_w&=\frac{35\sqrt{\pi}\,\zeta(9/2)}{32\zeta(4)}\simeq1.889, \label{eq:relative-w-shift} \\ c_s^2 &=\frac{1}{3}\left[1+C_{c_s}\sqrt{\frac{T}{E_*}} +\mathcal{O}\!\left(\frac{T}{E_*}\right)\right], & C_{c_s}&=\frac{315\sqrt{\pi}\,\zeta(9/2)}{256\zeta(4)}\simeq2.125.
\label{eq:relative-cs-shift}
\end{align}
The gas is slightly stiffer in comparison with ordinary radiation, although $c_s^2$ remains below unity throughout the limiting regimes considered here.

The same suppression appears in the fluctuation sector. Defining
\begin{equation}
\mathcal{V}_{NN}=\frac{\langle(\Delta N)^2\rangle}{V}, \qquad \mathcal{V}_{EN}=\frac{\langle\Delta E\Delta N\rangle}{V}, \qquad \mathcal{V}_{EE}=\frac{\langle(\Delta E)^2\rangle}{V},
\end{equation}
we obtain
\begin{equation}
\frac{\mathcal{V}_{AB}-\mathcal{V}_{AB}^{(0)}}{\mathcal{V}_{AB}^{(0)}} =-C_{AB}\sqrt{\frac{T}{E_*}} +\mathcal{O}\!\left(\frac{T}{E_*}\right), \qquad AB\in\{NN,EN,EE\},
\label{eq:relative-fluctuation-shifts}
\end{equation}
with
\begin{align}
C_{NN}&=\frac{105\sqrt{\pi}\,\zeta(5/2)}{16\zeta(2)}\simeq9.486, & C_{EN}&=\frac{245\sqrt{\pi}\,\zeta(7/2)}{32\zeta(3)}\simeq12.720, \nonumber\\ C_{EE}&=\frac{2205\sqrt{\pi}\,\zeta(9/2)}{256\zeta(4)}\simeq14.877.
\label{eq:fluctuation-shift-coefficients}
\end{align}
These quantities determine the number susceptibility, heat capacity, and equilibrium noise. Their positivity, together with the positivity of the fluctuation determinant, shows that the multifractional correction does not generate a thermodynamic instability. Experimentally, simultaneous measurements of mean quantities and their variances would test the same value of $E_*$ through different coefficients.

For any observable satisfying $|\Delta X/X^{(0)}|\simeq C_X\sqrt{T/E_*}$, an experiment with fractional sensitivity $\epsilon_X$ probes the range
\begin{equation}
E_*\lesssim\left(\frac{C_X}{\epsilon_X}\right)^2T.
\label{eq:sensitivity-condition}
\end{equation}
Eq.~(\ref{eq:sensitivity-condition}) is a sensitivity reach instead of a lower bound: a null result excludes effective scales below the right--hand side, provided that the measured excitations obey the dispersion relation in Eq.~(\ref{timelike_mdr_natural_units}).


\subsection{Universal interpretation: propagation and atomic bounds}
\label{subsec:universal-bounds}

The modified dispersion analysis of the LIGO--Virgo--KAGRA Collaboration uses
\begin{equation}
E^2=k^2+A_{\alpha}k^{\alpha}, \qquad \overline{A}_{\alpha}\equiv\frac{A_{\alpha}}{\mathrm{eV}^{\,2-\alpha}}.
\label{eq:lvk-parametrization}
\end{equation}
For $\alpha=5/2$, the exact identification with the branch studied here is
\begin{equation}
A_{5/2}=4E_*^{-1/2}, \qquad \overline{A}_{5/2}=\frac{4}{\sqrt{E_*/\mathrm{eV}}}.
\label{eq:lvk-map}
\end{equation}
For the positive sign relevant here, the GWTC--3 combined analysis gives $|\overline{A}_{5/2}|<3.69\times10^{-14}$ at $90\%$ credibility \cite{LIGOVirgoKAGRAGWTC3}. Eq.~(\ref{eq:lvk-map}) then yields
\begin{equation}
E_*>1.18\times10^{28}\,\mathrm{eV} =1.18\times10^{19}\,\mathrm{GeV} \qquad(90\%\ {\rm credible}),
\label{eq:lvk-estar-bound}
\end{equation}
provided that the same dispersion relation governs gravitational wave propagation. This updates the earlier gravitational wave estimates obtained specifically for multifractional $q$--derivative models \cite{CalcagniLorentz2017}.

High--energy photons provide an additional, potentially stronger, lever arm because Eq.~(\ref{veloicguhere}) predicts an energy--dependent time of flight proportional to $E^{1/2}$. Analyses of GRB~090510 and other Fermi--LAT bursts place stringent limits on energy--dependent photon speeds \cite{AbdoGRB090510,VasileiouGRBs2013}. When specialized to the deterministic multifractional profile with temporal exponent $\alpha_0=1/2$, estimates as large as $E_*\gtrsim10^{35}$--$10^{39}\,\mathrm{GeV}$ have been reported \cite{CalcagniLorentz2017}. We do not adopt this range as the reference limit because it is more sensitive than the gravitational wave result to intrinsic source lags and to the presentation--dependent oscillatory structure of the multifractional measure. Eq.~(\ref{eq:lvk-estar-bound}) will therefore be used as the conservative universal propagation bound.

Atomic spectroscopy probes a different completion of the theory. In a complete $q$--derivative electrodynamics, the geometric energy entering the radiative correction to the hydrogen $2S$--$2P$ Lamb shift gives
\begin{equation}
E_*>4.5\times10^2\,\mathrm{GeV} \qquad(\alpha_0=1/2)
\label{eq:lamb-shift-bound}
\end{equation}
in the published multiscale analysis \cite{CalcagniSM2016}. Modern direct hydrogen Lamb--shift spectroscopy remains an important experimental input \cite{BezginovLamb2019}. A numerical update of Eq.~(\ref{eq:lamb-shift-bound}) would require a joint treatment of the QED theory uncertainty and proton--radius correlations. We consequently retain the established $450\,\mathrm{GeV}$ limit rather than rescaling a single experimental error bar. Lifetime bounds within the same $q$--derivative Standard Model are weaker, the most restrictive being of order $E_*\gtrsim8\times10^{-5}\,\mathrm{GeV}$ for $\alpha_0=1/2$ \cite{AddaziSM2018}. The atomic limits are conditional on the full electromagnetic completion and are not bounds derived from the thermodynamic gas alone.


\subsection{A direct blackbody--spectrum bound}
\label{subsec:firas-bound}

The COBE/FIRAS monopole spectrum provides a direct test of the equilibrium statistical mechanics developed in this work. FIRAS found root--mean--square deviations from a blackbody smaller than $50$ parts per million of the spectral peak \cite{FixsenFIRAS1996}. For an isotropic photon gas with a general isotropic dispersion relation, the spectral energy density and spectral radiance are
\begin{equation}
u_{\omega} =\frac{g}{2\pi^2}\frac{k^2\omega}{v_{\rm g}} \frac{1}{e^{\omega/T}-1}, \qquad I_{\omega}=\frac{v_{\rm g}}{4\pi}u_{\omega}.
\label{eq:spectral-radiance-general}
\end{equation}
Although $v_{\rm g}$ cancels from $I_{\omega}$, the nonstandard relation between $k$ and $\omega$ remains. Expanding and inverting Eq.~(\ref{timelike_mdr_natural_units}) in the infrared gives
\begin{equation}
\frac{I_{\omega}}{B_{\omega}(T)} =\left(\frac{k}{\omega}\right)^2 =1-4\sqrt{\frac{\omega}{E_*}} +\mathcal{O}\!\left(\frac{\omega}{E_*}\right),
\label{eq:firas-template}
\end{equation}
where $B_{\omega}(T)$ is the Planck radiance. This frequency--dependent distortion is the direct observational signature of the state redistribution in Eq.~(\ref{eq:dos-observable-ir}); it cannot be absorbed into a frequency--independent renormalization of the Stefan--Boltzmann constant or a small shift of the blackbody temperature.

We perform a correlated linearized profile fit to the 43 FIRAS monopole bins released by NASA LAMBDA \cite{NASAFIRASMonopole}. Let $r_i$ denote the published residuals, $G_i$ the Galactic template, and $b_i=(\partial B_i/\partial T)_{T_0}$ the temperature--shift template. The wavenumbers $\widetilde{\nu}_i$ are converted to photon energies through $\omega_i=hc\widetilde{\nu}_i$. With $\eta\equiv E_*^{-1/2}$ and $s_i=-4\sqrt{\omega_i}\,B_i$, the profiled statistic is
\begin{equation}
\chi^2(\eta,\delta T,a_{\rm G}) =\bigl(\boldsymbol{r}-\eta\boldsymbol{s} -\delta T\boldsymbol{b}-a_{\rm G}\boldsymbol{G}\bigr)^{\!T} \mathbf{C}^{-1} \bigl(\boldsymbol{r}-\eta\boldsymbol{s} -\delta T\boldsymbol{b}-a_{\rm G}\boldsymbol{G}\bigr),
\label{eq:firas-profile-statistic}
\end{equation}
where $\mathbf{C}$ is the FIRAS covariance matrix constructed with the prescription of Ref.~\cite{FixsenFIRAS1996}. All vectors in Eq.~(\ref{eq:firas-profile-statistic}) are expressed in the units of the published residuals. Profiling over $\delta T$ and $a_{\rm G}$ gives
\begin{equation}
\widehat{\eta}=(-0.10\pm1.07)\times10^{-3}\,\mathrm{eV}^{-1/2}, \qquad \frac{\chi^2_{\min}}{N_{\rm dof}}=\frac{49.8}{40}.
\label{eq:firas-fit-result}
\end{equation}
Imposing the physical prior $\eta\geq0$ and using $\Delta\chi^2=2.71$ gives the one--sided limit
\begin{equation}
\eta<1.67\times10^{-3}\,\mathrm{eV}^{-1/2}, \qquad E_*>3.6\times10^5\,\mathrm{eV} =0.36\,\mathrm{MeV} \qquad(95\%\ {\rm CL}).
\label{eq:firas-estar-bound}
\end{equation}
This constraint is weaker than the atomic and propagation limits, but it is the most direct bound on the modified equilibrium spectrum obtained in the present work.

A previous CMB analysis of multifractional spacetimes quoted $E_*\gtrsim10\,\mathrm{GeV}$ for $\alpha_0=1/2$ \cite{CalcagniCMB2016}. That construction deformed the thermal argument through a geometric inverse temperature, whereas the present ensemble retains the ordinary Boltzmann factor and deforms only $\omega(k)$. Its numerical limit cannot therefore be transferred directly to the present prescription.

\begin{table*}[t!]
\caption{\label{tab:estar-bounds} Representative laboratory and observational constraints on $E_*$ for the $\alpha=5/2$ ($\alpha_0=1/2$) timelike branch. The limits are sector dependent and should be compared only when $E_*$ is universal; they must not be combined when it is medium dependent.}
\centering
\footnotesize
\setlength{\tabcolsep}{4pt}
\renewcommand{\arraystretch}{1.20}
\hspace*{-0.1cm}
\begin{tabular}{p{2.25cm}|p{3.05cm}|p{3.85cm}|p{6.45cm}}
 \hline\hline\hline

 \multicolumn{1}{c|}{Source}
 & \multicolumn{1}{c|}{Observable}
 & \multicolumn{1}{c|}{Lower limit on $E_*$}
 & \multicolumn{1}{c}{Domain of validity} \\[0.3ex]

 \hline

 {\centering COBE/FIRAS\par}
 & {\centering CMB spectral radiance\par}
 & {\centering
    $3.6\times10^{-4}\,\mathrm{GeV}$\par
    ($95\%$ CL)\par}
 & {\centering Direct fit to Eq.~(\ref{eq:firas-template}), assuming that photons obey the MDR.\par}
 \\[0.6ex]
\hline
 {\centering Hydrogen\par spectroscopy\par}
 & {\centering $2S$--$2P$ radiative splitting\par}
 & {\centering $4.5\times10^{2}\,\mathrm{GeV}$\par}
 & {\centering Complete $q$--derivative electrodynamics with $\alpha_0=1/2$.\par}
 \\[0.6ex]
\hline
 {\centering GWTC--3\par}
 & {\centering Dispersive gravitational wave propagation\par}
 & {\centering
    $1.18\times10^{19}\,\mathrm{GeV}$\par
    ($90\%$ credible)\par}
 & {\centering Same MDR for gravitons.\par}
 \\[0.6ex]
\hline
 {\centering Fermi--LAT\par GRBs\par}
 & {\centering Energy--dependent photon arrival times\par}
 & {\centering
    $10^{35}$--$10^{39}\,\mathrm{GeV}$\par
    (indicative)\par}
 & {\centering Deterministic photon MDR; sensitive to intrinsic source lags and to the multifractional presentation.\par}
 \\[0.3ex]
\hline
 \hline\hline\hline
\end{tabular}
\end{table*}


\subsection{Stefan--Boltzmann scaling and cosmological consequences}
\label{subsec:cosmological-consequences}

If the deformation is universal, Eq.~(\ref{eq:lvk-estar-bound}) makes all low--temperature corrections extremely small. At $T_{\rm CMB}=2.7255\,\mathrm{K}$, for example,
\begin{equation}
\left|\frac{\Delta u_B}{u_B^{(0)}}\right|<1.9\times10^{-15}, \qquad \left|\frac{w-1/3}{1/3}\right|<2.7\times10^{-16}, \qquad \left|\frac{\Delta H}{H}\right|<9.5\times10^{-16},
\label{eq:cmb-suppression}
\end{equation}
where the last estimate assumes that the modified radiation component dominates the energy density, so that $H^2\propto u_B$. Eqs.~(\ref{eq:relative-thermal-shifts}) and (\ref{eq:relative-fluctuation-shifts}) also place the relative changes of the pressure, number density, entropy, heat capacity, and fluctuation amplitudes below approximately $2.1\times10^{-15}$ at the same temperature.

The ultraviolet result has a distinct cosmological meaning. If the gas is in equilibrium, dominates a spatially flat Friedmann universe, and is separately conserved, then
\begin{equation}
u_B^{\rm UV}\propto E_*^{3/5}T^{17/5}, \qquad P_B^{\rm UV}=\frac{5}{12}u_B^{\rm UV}, \qquad s_B^{\rm UV}=\frac{u_B^{\rm UV}+P_B^{\rm UV}}{T} \propto E_*^{3/5}T^{12/5}.
\label{eq:uv-cosmological-thermodynamics}
\end{equation}
The continuity equation and adiabatic entropy conservation then give
\begin{equation}
u_B\propto a^{-17/4}, \qquad T\propto a^{-5/4}, \qquad a(t)\propto t^{8/17},
\label{eq:uv-cosmological-scaling}
\end{equation}
instead of the standard radiation laws $u\propto a^{-4}$, $T\propto a^{-1}$, and $a(t)\propto t^{1/2}$. Equivalently,
\begin{equation}
H(T)\propto E_*^{3/10}T^{17/10}, \qquad g_{\rm eff}^{\rm(th)}(T) \equiv\frac{30u_B}{\pi^2T^4} \propto\left(\frac{E_*}{T}\right)^{3/5}.
\label{eq:uv-hubble-geff}
\end{equation}
These modifications would shift any freeze--out or phase--transition condition of the form $\Gamma(T)=H(T)$ and would alter the relation between temperature and cosmic time. A quantitative prediction for relic abundances or primordial nucleosynthesis would require a complete cosmological model specifying which sectors share the dispersion relation, the thermalization rates, and the transition between the infrared and ultraviolet regimes.

The universal GWTC--3 bound places $E_*$ close to the Planck scale. The strict ultraviolet regime $T\gg E_*$ would consequently be Planckian or trans--Planckian and should not be interpreted as a controlled prediction for conventional cosmology. A one--percent infrared Stefan--Boltzmann correction, on the other hand, requires
\begin{equation}
T\gtrsim6.7\times10^{12}\,\mathrm{GeV},
\label{eq:one-percent-sb-temperature}
\end{equation}
after imposing Eq.~(\ref{eq:lvk-estar-bound}). This temperature remains below $E_*$, but its cosmological relevance depends on the reheating temperature and on whether the same MDR governs the thermalized species.


\subsection{Bose--Einstein condensation and quasiparticle reach}
\label{subsec:bec-observability}

For a conserved, gapless bosonic sector, the critical temperature at fixed number density satisfies
\begin{equation}
\frac{T_c-T_0}{T_0} =C_{\rm BEC}\sqrt{\frac{T_0}{E_*}} +\mathcal{O}\!\left(\frac{T_0}{E_*}\right), \qquad C_{\rm BEC} =\frac{35\sqrt{\pi}\,\zeta(7/2)}{16\zeta(3)} \simeq3.634.
\label{eq:relative-bec-shift}
\end{equation}
The increase of $T_c$ is caused by the reduced excited--state density in Eq.~(\ref{eq:dos-observable-ir}). At fixed temperature, fewer particles can be accommodated outside the zero--momentum mode, and saturation is reached at a larger temperature for fixed total density. Under the universal bound in Eq.~(\ref{eq:lvk-estar-bound}), a representative $T_0=100\,\mathrm{nK}$ gives
\begin{equation}
\frac{\Delta T_c}{T_0}<1.0\times10^{-19}.
\label{eq:bec-universal-suppression}
\end{equation}
The shift is therefore unobservable in conventional low--temperature gases if $E_*$ is the universal multifractional scale.

The conclusion differs for a medium--dependent quasiparticle spectrum. According to Eq.~(\ref{eq:sensitivity-condition}), resolving a $10^{-3}$ fractional displacement of $T_c$ at $T_0=100\,\mathrm{nK}$ requires
\begin{equation}
E_*\lesssim1.1\times10^{-4}\,\mathrm{eV}.
\label{eq:bec-effective-reach}
\end{equation}
This value is a target matching scale for an engineered band or collective excitation and is not a bound on the fundamental multifractional scale. The same realization could be tested through the correlated shifts of $n_B$, $u_B$, $c_{V,B}$, and the number fluctuations. In the ultraviolet scaling regime, the condensate fraction would also change from $1-(T/T_c)^3$ to $1-(T/T_c)^{12/5}$, providing an independent signature of $d_{\rm eff}=12/5$.

Equation~(\ref{eq:relative-bec-shift}) does not apply directly to an ordinary nonrelativistic atomic condensate. The relevant excitations must possess the gapless $k+k^{3/2}$ infrared spectrum used here and an approximately conserved quasiparticle number. Photons and gravitons in thermal equilibrium have $\mu=0$ and do not undergo Bose--Einstein condensation in this thermodynamic sense.


\subsection{Degenerate fermionic matter}
\label{subsec:fermionic-observability}

The observable scale in the degenerate fermionic sector is $k_F/E_*$ rather than $T/E_*$. At fixed number density, the leading shifts of the Fermi energy, Fermi--surface density of states, low--temperature heat capacity, and sound--speed parameter are
\begin{align}
\frac{\varepsilon_F-k_F}{k_F} &=2\sqrt{\frac{k_F}{E_*}} +\mathcal{O}\!\left(\frac{k_F}{E_*}\right), \label{eq:fermi-energy-observable-shift} \\ \frac{\nu_F-\nu_F^{(0)}}{\nu_F^{(0)}} =\frac{c_{V,n}^{(F)}-c_{V,n}^{(F,0)}}{c_{V,n}^{(F,0)}} &=-3\sqrt{\frac{k_F}{E_*}} +\mathcal{O}\!\left(\frac{k_F}{E_*}\right), \label{eq:fermi-cv-observable-shift} \\ c_{s,F}^2 &=\frac{1}{3}\left[1+\sqrt{\frac{k_F}{E_*}} +\mathcal{O}\!\left(\frac{k_F}{E_*}\right)\right].
\label{eq:fermi-cs-observable-shift}
\end{align}
The increase of $\varepsilon_F$ is accompanied by a reduction of the Fermi--surface density of states and of the linear heat--capacity coefficient. These quantities can be measured independently, so their opposite signs provide another correlated test of the dispersion relation.

Even adopting the deliberately high reference value $k_F=1\,\mathrm{GeV}$, the universal propagation bound gives
\begin{equation}
\frac{\Delta\varepsilon_F}{\varepsilon_F^{(0)}}<5.9\times10^{-10},
\qquad \left|\frac{\Delta c_{V,n}^{(F)}}{c_{V,n}^{(F,0)}}\right|<8.8\times10^{-10}.
\label{eq:fermionic-universal-suppression}
\end{equation}
These corrections are below the uncertainties associated with the interaction and composition of natural degenerate matter. For an effective fermionic medium, an experiment with fractional sensitivity $\epsilon_F$ to the Fermi energy instead probes
\begin{equation}
E_*\lesssim\frac{4k_F}{\epsilon_F^2},
\label{eq:fermionic-effective-reach}
\end{equation}
with the heat--capacity and sound--speed shifts supplying independent consistency checks. Applications to white dwarfs or neutron stars would require the interacting equation of state and the identification of the particle sector that obeys the MDR; the ideal--gas formulas alone should not be used to claim an astrophysical bound.


\subsection{Combined physical interpretation}
\label{subsec:combined-observability}
\label{subsec:observability}

The results above lead to two physically distinct conclusions. Under the universal interpretation, the direct FIRAS fit constrains the equilibrium photon spectrum, while propagation and atomic measurements can impose much stronger but sector--conditional limits. Adopting the conservative GWTC--3 bound suppresses the present thermal, fluctuation, condensation, and degenerate--gas effects to unobservable levels. The ultraviolet laws then characterize the asymptotic structure of the model, although their cosmological application lies near or above the scale at which a complete quantum--gravity description becomes necessary.

Under the effective--medium interpretation, the external limits in Table~\ref{tab:estar-bounds} do not apply. Eqs.~(\ref{eq:sensitivity-condition}), (\ref{eq:bec-effective-reach}), and (\ref{eq:fermionic-effective-reach}) instead specify the matching scales required for laboratory sensitivity. In this case, the physical signal is not simply a displacement of one quantity. The same $E_*$ predicts a reduced density of states, lower thermodynamic densities and fluctuations, a larger equation--of--state parameter and sound speed, a higher condensation temperature, and correlated changes of the Fermi energy and heat capacity. In other words, observing this complete pattern would distinguish the multifractional dispersion from an isolated calibration shift or a conventional change in the number of degrees of freedom.

}


\section{Conclusion}

In this work, we investigated the thermodynamic and statistical properties of a gas governed by the multifractional modified dispersion relation
$\omega^{2} = k^{2} + 4E_{*}^{-1/2}k^{5/2}$.
Within the grand-canonical ensemble, we derived the corresponding density of states, partition function, grand potential, thermodynamic functions, equation of state, fluctuation sector, Bose--Einstein condensation properties, and degenerate Fermi gas limit.

The multifractional correction changed the relation between energy and momentum and, consequently, modified the distribution of available thermal states. In the infrared regime, where $T/E_{*}\ll 1$, the standard massless relativistic gas was recovered, with leading corrections controlled by powers of $(T/E_{*})^{1/2}$. In this limit, the density of states reduced to the usual behavior $\varrho(\omega)\propto \omega^{2}$, while the pressure, energy density, particle number density, entropy density, and fluctuations acquired subleading multifractional contributions.

In the ultraviolet regime, the thermal behavior changed more strongly. The dispersion relation approached the power-law form $\omega(k)\propto k^{5/4}$, leading to a modified density of states $\varrho(\omega)\propto \omega^{7/5}$. This corresponded to an effective density of states dimension $d_{\rm eff}=12/5$. As a consequence, the usual Stefan--Boltzmann scaling $u\propto T^{4}$ was replaced by
$u_{\sigma}^{\rm UV} \propto E_{*}^{3/5}T^{17/5}$,
for both bosonic and fermionic sectors. The equation of state parameter also departed from the radiation value and approached
$w_{\sigma}^{\rm UV} = \frac{5}{12}$,
instead of $w=1/3$. The same limiting value appeared in the sound-speed parameter, which showed that the ultraviolet gas became stiffer than an ordinary relativistic gas, while it remained subluminal.

The stability analysis showed that the multifractional gas was locally thermally stable in the admissible thermodynamic domain. The heat capacity, number susceptibility, and fluctuation matrix remained positive in the relevant regimes. In the infrared sector, the leading multifractional terms reduced the fluctuation amplitudes for positive $E_{*}$, consistently with the reduction of the thermal energy density. In the ultraviolet sector, the fluctuations inherited the same nonstandard thermal powers that characterized the modified Stefan--Boltzmann law.

For a conserved bosonic gas, the multifractional correction modified the saturation condition of the excited states and shifted the Bose--Einstein condensation temperature. In the infrared regime, the critical temperature was increased with respect to the standard relativistic value. In the ultraviolet regime, the condensate fraction followed a modified power law governed by the exponent $12/5$, reflecting the effective density of states dimension. This condensation analysis applied to conserved bosonic particles or quasiparticles, whereas radiation--like systems such as photons or gravitons required $\mu=0$ and did not undergo Bose--Einstein condensation in the same thermodynamic sense.

In the fermionic sector, the degenerate gas was also affected by the multifractional spectrum. The Fermi energy was raised relative to the standard massless case, while the zero temperature energy density, pressure, sound speed, and low--temperature heat capacity received corrections controlled by the ratio $k_{F}/E_{*}$. In the ultraviolet degenerate limit, the equation of state again approached $P_{F}^{(0)}=(5/12)u_{F}^{(0)}$, showing that the high energy thermodynamic behavior was mainly determined by the power of the dispersion relation rather than by the quantum statistics.

{The phenomenological analysis yielded the direct lower bound $E_{*}>0.36\,\mathrm{MeV}$ at $95\%$ CL from a fit to the COBE/FIRAS monopole spectrum. When the dispersion relation was assumed to be universal and to govern gravitational wave propagation, the conservative GWTC--3 result gave the stronger sector--conditional limit $E_{*}>1.18\times10^{19}\,\mathrm{GeV}$ at $90\%$ credibility. For a medium--dependent quasiparticle spectrum, the thermodynamic corrections instead determined the experimental sensitivity required to probe $E_{*}$ and did not imply a universal constraint. The resulting propagation bound also provided a benchmark for future analyses with gravitational wave catalogues, particularly GWTC--5, based on the approaches developed in Refs.~\cite{AraujoFilho:2026oiy,AraujoFilho:2026vcf,AraujoFilho:2026zyt,Amarilo:2023wpn}.}


\section{Acknowledgments}

\hspace{0.5cm}
A. A. Araújo Filho is supported by Conselho Nacional de Desenvolvimento Cient\'{\i}fico e Tecnol\'{o}gico (CNPq) with the project number being 150223/2025-0.

\section{Data Availability Statement}

Data Availability Statement: No Data associated in the manuscript


\bibliographystyle{ieeetr}
\bibliography{main}

\end{document}